\documentclass[11pt,a4paper]{amsart}
\usepackage{lscape,amsmath,amssymb,setspace,tikz,longtable,url,pgfplots,graphics}
\pgfplotsset{compat=newest}
\usepackage{thmtools}
\usepackage{tasks}
\usepackage{pgfplots}
\usepackage{float}
\usepackage[width=1.00\textwidth]{caption}
\usetikzlibrary{arrows}
\usepgfplotslibrary{fillbetween}
\declaretheoremstyle[
spaceabove=7pt, spacebelow=7pt,
headfont=\normalfont\bf,
notefont=\normalfont, notebraces={(}{)},
bodyfont=\normalfont,
]{break}
\theoremstyle{break}

\newtheorem{rem}{Remark}

\newtheorem{ex}{Example}

\begin{document}
	
\title[Inequality measurements]{Inequality measures:
  The Kolkata index in comparison with other measures}

\author[Banerjee]{Suchismita Banerjee} \address{Economic Research
Unit, Indian Statistical Institute, Kolkata, India}
\email{suchib.1993@gmail.com}
\author[Charkrabarti]{Bikas K. Chakrabarti} \address{Saha Institute of Nuclear Physics, Kolkata, \and Economic
Research Unit, 	\vskip0.5ex Indian Statistical Institute, Kolkata, India}
\email{bikask.chakrabarti@saha.ac.in}
\author[Mitra]{Manipushpak Mitra} \address{Economics Research Unit,
Indian Statistical Institute, Kolkata, India}\email{mmitra@isical.ac.in}
\author[Mutuswami]{Suresh Mutuswami} \address{School of Business,
University of Leicester, Leicester, United Kingdom}
\email{smutuswami@gmail.com}

\maketitle

\begin{abstract}
  \noindent We provide a survey of the \emph{Kolkata index} of social inequality
  (\cite{GC2014}, \cite{CG2017}, \cite{BCMM2020}), focusing in
  particular on income inequality. Based on the
      observation that inequality functions (such as the Lorenz
      function), giving the measures of income or wealth against that
      of the population, to be generally nonlinear, we show that the
      fixed point (like Kolkata index $k$) of such a nonlinear
      function (or related, like the complementary Lorenz function)
      offer better measure of inequality than the average quantities
      (like Gini index).  Indeed the Kolkata index can be viewed as a
      generalized Hirsch index for a normalized inequality function
      and gives the fraction $k$ of the total wealth possessed by the
      rich $1-k$ fraction of the population. We analyze the
  structures of the inequality indices for both continuous and
  discrete income distributions. We also compare the Kolkata index to
  some other measures like the Gini coefficient and the Pietra index.
  Lastly, we provide some empirical studies which illustrate the
  differences between the Kolkata index and the Gini coefficient.
  
\vspace{0.5cm}  
	
\end{abstract}  

\maketitle

\section{Introduction} 
\noindent Inequality in a society can broadly be categorized as \emph{inequality
  of condition} or \emph{inequality of opportunity}.  The former
refers to disparities in the current status of individuals, whether
this be income, wealth or their ownership of different goods and
services.  The latter refers to disparities in the future potential of
individuals.  Typically, inequality of opportunity is inferred
indirectly through its effects like education level and quality,
health status and treament by the justice system.  Though the two
types of inequality are interrelated, we are interested in the former
type only in this survey.  % Inequality
% of condition has serious social consequences like health, education,
% incidence of crime and violence (see \cite{Deaton2001}).
Therefore, in what follows, the term `inequality' will refer
exclusively to inequality of condition.  % We will also, for the most
% part, confine ourselves to the case of income inequality.

We focus here on one aspect of inequality, viz., the measurement of
inequality.
% Inequality measures are often employed to judge the distributional
% effects of a particular economic policy or to simply evaluate a
% particular income distribution.
Measuring inequality is important for answering a wide range of
questions.  For instance: is the income distribution more equal than
what it was in the past? Are underdeveloped countries characterized by
greater inequality than developed countries? Do taxes or other kinds
of policy interventions lead to greater equality in the distribution
of income or wealth?  Since the way inequality is measured also
determines how the above questions (among others) are answered, a
rigorous discussion of the measurement of inequality is necessary (see
e.g., \cite{SRC2015} and \cite{Jenkins2011}).

A tool that is indispensable in measuring income and wealth inequality
is the \emph{Lorenz function} and its graphical representation, the
\emph{Lorenz curve} (see \cite{Lo1905}).  The Lorenz curve plots the
percentage of total income earned by various portions of the
population when the population is ordered by the size of their
incomes. The Lorenz curve is typically depicted as a curve in the unit
square with end points at $(0, 0)$ and $(1,1)$ (see Figure
\ref{fig:Lorenz curve}).\footnote{The end points are clear since none
  of the population posseses none of the income while the entire
  population possesses all the income.}  The $45^\circ$ line is the
\emph{line of perfect equality} representing a situation where all
individuals have the same income.

The Lorenz curve can be used, in a limited way, as a measure of
inequality.  Since the $45^\circ$ line is the line of perfect
equality, we can say that the ``closer'' a Lorenz curve is to the
$45^\circ$ line, the more equal is the income distribution.
Unfortunately, this does not get us very far because Lorenz curves can
intersect and hence, the Lorenz curves cannot be ranked unambiguously
using the above criterion (see \cite{Aa2000}). We have more to say on
this point in Section
2. % This possibility of ambiguity associated with
% comparing Lorenz curves is also established in Section 2 of this
% survey using simple examples. Therefore, Lorenz function and its
% associated Lorenz curve, though indispensable, cannot be used as a
% measure of inequality.

The existing literature sees two approaches to deal with the problem
of intersecting Lorenz curves. The first is to consider ranking
criterion that are `weaker' than this dominance criterion meaningful
only for non-intersecting Lorenz curves (see \cite{Aa2000},
\cite{DL1988}, \cite{Lam1993}, \cite{SF1987} and \cite{ZOLI1999}). The
pioneering work in this approach is \cite{Da1920} which suggested that
there is an underlying notion of social welfare associated with any
measure of income inequality.  It is this concept with which we should
be concerned. Furthermore, we should approach the question by
considering directly the form of the social welfare function to be
employed (see \cite{Atkinson1970}). This is a normative approach and
is meaningful when we want to obtain a ranking of income distributions
in order to infer something from the social welfare angle like whether
``post-tax income is more equally distributed than pre-tax income''.

The second approach is to develop summary measures of inequality using
the Lorenz functions (see \cite{Aa2000} for details).  Here, each
Lorenz function is associated with a real number and these numbers are
used to compare inequality across different income distributions. This
is a descriptive approach where we quantify the difference in
inequality between pairs of distributions (see \cite{Atkinson1970}).
 
An index of income inequality is therefore a scalar measure of
interpersonal income differences within a given population. High
income inequality means concentration of high incomes in the hands of
few and is likely to compress the size of the middle class. A large
and rich middle class contributes significantly to the well-being of a
society in many ways. In particular, a large and rich middle class
contributes in terms of high economic growth, better health status,
higher education level, a sizeable contribution to the country's tax
revenue and a better infrastructure, and more social cohesion
resulting from fellow feeling. A society characterized with a small
middle class and more persons away from the middle income group may
lead to a strained relationship between the subgroups on the two sides
of the middle class which can generate unrest (see
\cite{SRC2015}). Hence, the need for identifying the magnitude of
income inequality through different indices is of prime importance.

Except for the unique case of equality, where the Lorenz curve is
trivially linear, the Lorenz function is typically nonlinear and it
accommodates the essential features of the inequalities
involved. However, most of the common inequality indices formulated
and used so far studies some of the `average' properties of the Lorenz
function. On the other hand, the established observations in
statistical physics, for example in developing the Renormalization
Group theory of phase transitions (see e.g., \cite{Fisher1998}) or the
chaos theory (see e.g., \cite{Feigenbaum1983}), strongly indicated the
richness of the (nontrivial) fixed point structure (and also of the
eigen vectors and eigen values for the linearized function near that
fixed point) of such non-linear functions to comprehend the physical
and mathematical process represented by such nonlinear functions. We
noted earlier (see \cite{GC2014}) that, while the Lorenz function has
got trivial fixed points, a complementary Lorenz function has a
non-trivial point corresponding to an inequality index called the
Kolkata index, having several intriguing and useful properties.
 
Our primary focus in this survey will be on the Kolkata index as a
measure of inequality. The \emph{Kolkata index}, first introduced by
\cite{GC2014} and later analyzed in \cite{CG2017} and in
\cite{BCMM2020}, is that proportion $k$ of the population such that
the proportion of income that we can associate with $k$ is $(1-k)$.
Since no single summary statistic can reflect all aspects of
inequality exhibited by the Lorenz curve, the importance of using
alternative measures of inequality is universally acknowledged (see
\cite{Aa2000}). We would also discuss two popular indices namely, the
\emph{Gini coefficient or index} (see
\cite{Gi1912}) and the \emph{Pietra index} (see \cite{Pi1915}). The
Gini index is the ratio of the area between the $45^\circ$ line and
the Lorenz curve to the total area under the $45^\circ$
line. Equivalently, the Gini index is twice the area between the
Lorenz curve and the line of perfect equality. The Pietra index is the
maximum value of the gap between the $45^\circ$ line and the Lorenz
curve (also see \cite{ES2010}).

    In section $2$, we discuss the fundamentals of
    Lorenz and complementary Lorenz functions, along with some
    examples extending from continuous to discrete wealth
    distributions. In section $3$, we define the Kolkata index
    ($k$-index) and show some example calculation of the $k$-index for
    continuous wealth distributions. We also demonstrate an algorithm
    for calculating the $k$-index for discrete wealth distribution. We
    conclude the section by comparing the $k$-index with various other
    indices. In section $4$ and $5$, we continue this comparison based
    on rich-poor disparity. In section $6$, we measure the $k$-index
    from real society data. Section $7$ summarizes and concludes this
    work.

\section{Lorenz function and the complementary Lorenz function}
 \noindent Let $F$ be the distribution function of a non-negative random variable
$X$ which represents the income distribution in a society. The
left-inverse of $F$ is defined as
$F^{-1}(q) = \inf_{x}\{x\in X|F(x) \geq q\}$. As long as the mean
income $\mu = \int_{0}^{\infty}xdF(x)$ is finite, we obtain an
alternative representation of the mean:
$\mu = \int_{0}^{1}F^{-1}(q)dq$.  The function associated with the
Lorenz curve is the \emph{Lorenz function}, defined as
$L_F(p) =(1/\mu)\int_{0}^{p}F^{-1}(q)dq$. The Lorenz function gives
the proportion of total income earned by the bottom $100p\%$ of the
population for every given $p\in [0,1]$. The advantage of this
definition of Lorenz function due to \cite{Ga1971} is that it can be
applied to income distributions with both discrete and continuous
random variables. The Lorenz function thus defined has the following
properties: (i) $L_F(p)$ is continuous, non-decreasing and convex in
$p\in (0,1)$ and , (ii) $L_F(0)=0$, $L_F(1)=1$ and $L_F(p)\leq p$ for
all $p\in (0,1)$. Moreover, if there exists $p\in (0,1)$ such that
$L_F(p)=p$, then for all $p\in [0,1]$, $L_F(p)=p$. If the Lorenz
function $L_F(p)$ is differentiable in the open interval $(0,1)$, then
the slope of the Lorenz function at any $p\in (0,1)$ is given by
$F^{-1}(p)/\mu$. Let $M_F$ be the median as a percentage of the
mean. Then $M_F$ is given by the slope of the Lorenz curve at
$p =1/2$, that is, $M_F=F^{-1}(1/2)/\mu$. Since many real life
distributions of incomes are skewed to the right, the mean often
exceeds the median so that $M_F<1$. The \emph{complementary Lorenz
  function} is defined as $\hat{L}_F(p) = 1 - L_F(p)$ . It measures
the proportion of the total income earned by the top $100(1 - p)\%$ of
the population. Therefore,

\begin{equation}
  \hat{L}_F(p) := 1 - L_F(p) = 1 - \frac{\int\limits_{0}^{p}F^{-1}(q)dq}{\mu} = \frac{\int\limits_{p}^{1}F^{-1}(q)dq}{\mu}.
\end{equation}

 It easily follows that $\hat{L}_F(0) = 1, \hat{L}_F(1) = 0$, and
$0 \leq \hat{L}_F(p) \leq 1$ for $p \in (0, 1)$.  Furthermore,
$\hat{L}_F(p)$ is continuous, non-increasing and concave for
$p\in (0, 1)$.

\begin{figure}[h]
	\begin{center}
		\begin{tikzpicture}[scale=6]
                  \draw (1, 1)--(0, 1)--(0, 0); \draw [fill=yellow]
                  (0, 0)..controls(0.75, 0.25)..(1, 1)--(0, 0); \draw
                  [fill=pink] (0, 0)..controls(0.75, 0.25)..(1, 1) --
                  (1, 0) -- (0, 0); \draw[green,thick] (0, 0)--(1,
                  0)--(1, 1); \draw
                  [orange,thick](0,1)..controls(0.75, 0.75)..(1, 0);
                  \draw [blue,thick] (0, 0)--(1, 1) node[pos=0,below
                  left]{(0, 0)}node [pos=1, right]{(1, 1)}; \draw (0,
                  1)--(1, 0) node [pos=0,left] {(0, 1)} node
                  [pos=1,below ] {(1, 0)}; \draw [red,thick](0,
                  0)..controls(0.75, 0.25)..(1, 1); \draw [->]
                  (0.935,0.81)--(1.1,0.81)node [pos=1,right] {Lorenz
                    curve};
                  \draw[->](0.55,0.55)--(1.1,0.55)node[pos=1,right]{perfect
                    equality line};
                  \draw[->](1,0.3)--(1.1,0.3)node[pos=1,right]{perfect
                    inequality line}; \draw [->]
                  (0.95,0.15)--(1.1,0.15)node [pos=1,right]
                  {complementary Lorenz curve}; \draw
                  [dashed](0.688,0)--(0.688,0.688)node [pos=0,below ]
                  {Q}; \draw [black,thick] (0.5,0.5)--(0.685,0.316)
                  node[pos=0,left]{O} node[pos=1,right]{P};
		\end{tikzpicture}
		\caption {The Lorenz and the
                    complementary Lorenz curves.
                    % related
                    % quantities.
                    % \textcolor{red}{Red line}: corresponds
                    % to a general Lorenz Curve ($L_F(p)$),
                    % \textcolor{orange}{Orange line}: defines the
                    % Complementary Lorenz Curve,
                    % $\hat{L}_F(p)=1-L(p)$,
                    % \textcolor{blue}{Blue line}: represents line of
                    % perfect equality whereas \textcolor{green}{Green
                    % line}: represents line of perfect
                    % inequality.
                    $Q$ is the $k$-index of the Lorenz
                    curve. $\overline{OP}$ represents the maximum
                    distance between the perfect equality line and the
                    Lorenz curve.}
		\label{fig:Lorenz curve}
	\end{center}
\end{figure}
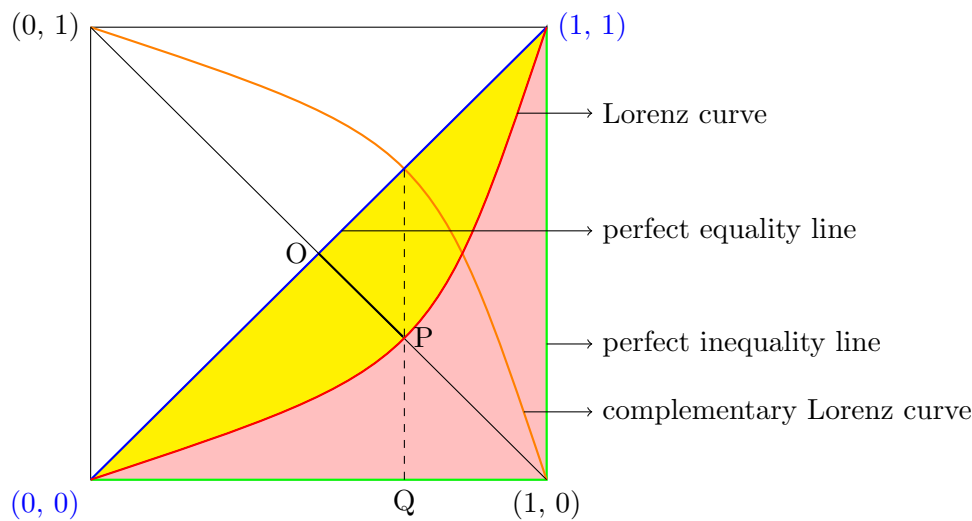

 Consider any egalitarian income distribution $F_e$ where all agents
earn a common positive income so that the associated Lorenz function
is $L_{F_e}(p)=p$ for all $p\in (0,1)$. Thus, we have a case of
perfect equality where every $p$\% of the population enjoys $p$\% of
the total income and the Lorenz curve coincides with the diagonal line
of perfect equality. In reality, we do not find any society where all
individuals have equal income. For all other income distributions the
Lorenz curve will lie below the egalitarian line, that is below the
Lorenz curve associated with the Lorenz function $L_{F_e}(.)$ for the
egalitarian income distribution $F_e$. Similarly, we also do not find
a society where one person has all the income, that is, an income
distribution $F_I$ such that $L_{F_I}(p)=0$ for all $p\in
(0,1)$. Specifically, with complete inequality associated with the
income distribution $F_I$, which is characterized by the situation
where only one agent has positive income and all other persons have
zero income, the Lorenz curve will run through the horizontal axis
until we reach the richest person and then it rises perpendicularly
(see Figure \ref{fig:Lorenz curve}). Hence, for any realistic income
distribution of a society, Lorenz curve always lie in between the
perfect equality line and the perfect inequality line. The Lorenz
curve is quite useful because it shows graphically how the actual
distribution of incomes differs not only from the perfect equality
line associated with the egalitarian income distribution $F_e$ but
also from the perfect inequality line associated with the income
distribution $F_I$. The Lorenz curve, complimentary Lorenz curve,
perfect equality and perfect inequality lines are shown in the Figure
\ref{fig:Lorenz curve} below, where we plot the fraction of population
from poorest to richest on the horizontal axis and the fraction of
associated income on the vertical axis.

We provide some simple examples of Lorenz functions for which the
associated income distribution is a continuous random variable.

\begin{itemize}
\item \emph{Uniform distribution:} Consider a society where the income
  distribution is uniform on some compact interval $[a,b]$ with
  $0\leq a<b<\infty$ so that the probability density function is
  $f_u(x)=1/(b-a)$ and the distribution function is
  $F_u(x)=(x-a)/(b-a)$ for every $x\in [a,b]$. Since $\mu_u=(a+b)/2$
  and $F^{-1}_{u}(q)=a+(b-a)q$, we get
  
\begin{equation*}
  L_{F_u}(p)=\frac{1}{\mu_u}\int_{0}^p\left\{a+(b-a)q\right\}dq=p\left[1-\frac{(b-a)}{(a+b)}(1-p)\right]. 
\end{equation*}

 Observe that if $a=0$, then we have $L_{F_{\bar{u}}}(p)=p^2$.
\item \emph{Exponential distribution:} Suppose the income distribution
  is exponential so that the probability density function is given by
  $f_E(x)=\lambda e^{-\lambda x}$ with $\lambda>0$ and the
  distribution function is $F_E(x)=1-e^{-\lambda x}$ for any
  $x\geq 0$. In this case $\mu_E=1/\lambda$ and
  $F^{-1}_E(q)=-(1/\lambda)\ln (1-q)$ implying

  \begin{equation*}
    L_{F_E}(p)=\int_0^p-\ln(1-q)dq=-\int_{t=1-p}^{t=1}\ln(t)dt=p-(1-p)\ln\left(\frac{1}{1-p}\right). 
\end{equation*}

\item \emph{Pareto distribution:} Consider a society where the income
  distribution is Pareto so that the density function is
  $f_{P,\alpha}(x)=\alpha(m)^{\alpha}/(x)^{\alpha+1}$ and the
  distribution function is $F_{P,\alpha}(x)=1-(m/x)^\alpha$ where
  $m>0$ is the minimum income, $\alpha>1$ and the density and
  distribution functions are defined for all $x\geq m$. In this case
  $\mu_P=\alpha m/(\alpha-1)$ and
  $F^{-1}_{P,\alpha}(q)=m(1-q)^{-(1/\alpha)}$ implying
\begin{equation}\label{Paretodis}
L_{F_{P,\alpha}}(p)=\frac{(\alpha-1)}{\alpha}\int_0^p(1-q)^{-\frac{1}{\alpha}}dq=\left[t^{\frac{(\alpha-1)}{\alpha}}\right]_{t=1-p}^{t=1}=1-(1-p)^{1-\frac{1}{\alpha}}. 
\end{equation}  
\end{itemize}

Hence, if the income distribution is a continuous random variable $F$,
one can calculate the Lorenz function $L_F(p)$ and, using
$\hat{L}_F(p)=1-L_F(p)$, we can easily calculate the associated
complementary Lorenz function as well.

\begin{ex}\label{drv}
   \emph{Discrete random variable}. To understand the procedure for
  getting the Lorenz function for income distribution given by
  discrete random variables, consider an economy with $G$ groups of
  people where each group $g\in \{1,\ldots,G\}$ has a total of
  $n_g\geq 1$ people with each person within this group having the
  same income of $x_g$ and also assume that $0\leq x_1<\ldots
  <x_G$. Define the total population as $N:=\sum_{g\in G}n_g$ and the
  total income of the economy as $M:=\sum_{g\in G}n_gx_g$ so that the
  mean income for this society is $\mu_G=M/N$. This income
  distribution is a discrete random variable $X=\{x_1,\ldots,x_G\}$
  such that the probability mass function is given by $f_G(x_g)=n_g/N$
  for all $g\in \{1,\ldots,G\}$ and the distribution function is given
  by
	\begin{equation*}
          F_G(x) = \left\{\begin{array}{cl} 0 & \mbox{ if
                                                $x\in [0,x_1)$,} \\ 
                            \frac{\left\{\sum\limits_{t=1}^gn_t\right\}}{N} & \mbox{if $x\in [x_g,x_{g+1})$ for any given $g\in \{1,\ldots,G-1\}$,}\\
                            1 & \mbox{if $x\geq x_G$.}
	\end{array} \right.
    \end{equation*}

    For each $g\in \{1,\ldots,G\}$, define $N(g):=\sum_{t=1}^gn_t/N$,
    $N(0):=0$, $M(g):=\sum_{t=1}^gn_tx_t/M$ and $M(0):=0$. For any
    given $g\in \{1,\ldots,G\}$ and any
    $q_g\in \left(N(g-1),N(g)\right]$, one can easily verify that
    $F^{-1}_G(q_g)=x_g$. Hence, using the Lorenz function formula we
    have the following: For any given $g\in \{1,\ldots,G\}$ and any
    $p_g\in \left(N(g-1),N(g)\right]$,

    \begin{equation}\label{examplezeroLorenz}
	L_{F_G}(p_g)=M(g-1)+\left(p_g-N(g-1)\right)\left(\frac{Nx_g}{M}\right).
	\end{equation}
	The following observations are helpful in this context.

	\begin{enumerate}
        \item The Lorenz function $L_{F_G}(p)$ is piecewise linear
          and, for each $g\in \{1,\ldots,G-1\}$, the point
          $(N(g),L_G(N(g))=M(g))$ on the coordinate plane of the graph
          of the Lorenz curve is a kink point.
        \item If $G=1$ so that $M=Nx_1$, $N(1)=1$, then from
          (\ref{examplezeroLorenz}) we get
          $L_{F_1}(p)=M(0)+\left(p-N(0)\right)(Nx_1/Nx_1)=p$ for all
          $p\in (N(0),N(1)]$, that is, Lorenz curve is associated with
          the egalitarian distribution and we have
          $L_{F_1}(p)=L_{F_e}(p)=p$ for all $p\in (0,1)$.
	\end{enumerate}    
\end{ex}

%  \begin{ex}\label{GParetoLaw}   Consider the distribution where a
% fraction $f$ of the population receives income $m_0$ and the remaining
% fraction $(1-f)$ recieves an income according to a Pareto law with
% origin $m_0$ and parameter $\alpha>1$.This distribution is a
% combination of a guaranteed minimum income $m_0$ with a Pareto
% distribution and was used by Gastwirth \cite{Ga1971}. The distribution
% function is given by
	
%  \begin{equation*}  F_m(x) = \left\{\begin{array}{cl} 0 & \mbox{ if
%  $x<m_0$,} \\  f & \mbox{if $x=m_0$,}\\ %
% 1-(1-f)\left(\frac{m_0}{x}\right)^{\alpha} & \mbox{if $m_0<x<\infty$.}
%  \end{array} \right.
%  \end{equation*}Here $\mu_m=m_0(\alpha-f)/(\alpha-1)$. The
% left-inverse of the distribution function $F_m$ is given by
%  \begin{equation*}  F^{-1}_m(t) = \left\{\begin{array}{cl} m_0 &
% \mbox{ if % $t\in (0,f]$,}\\ 
% m_0\left(\frac{1-f}{1-p}\right)^{1/{\alpha}} & \mbox{if $t\in (f,1)$.}
%  \end{array} \right.
%  \end{equation*}Applying the Lorenz function formula we get
%  \begin{equation*}  L_m(p) = \left\{\begin{array}{cl}
% \left(\frac{\alpha-1}{\alpha-f}\right)p & \mbox{ if  $p\in (0,f]$,}\\

% 1-\frac{\alpha(1-f)^{1/{\alpha}}}{(\alpha-f)}(1-p)^{(\alpha-1)/\alpha}
% & \mbox{if $p\in (f,1)$.}
%  \end{array} \right.
%  \end{equation*}Observe that if $f=0$, then $L_m(p)$ reduces to the
% Lorenz function associated with the Pareto distribution $L_P(p)$ given
% in (\ref{Paretodis}).
%  \end{ex}

\subsection{The Lorenz function as a measure of inequality} The Lorenz curve allows us to rank distributions according to
inequality and say that the country with Lorenz curve closer to the
perfect equality line has less inequality than the country with Lorenz
curve further away.  Consider two societies with income distributions
given by the distribution functions $F_a$ and $F_b$. If it so happens
that $L_{F_a}(p)\leq L_{F_b}(p)$ for all $p\in [0,1]$, then clearly,
the society with income distribution $F_a$ is more unequal compared to
the society having the income distribution $F_b$ since for every
$p\in (0,1)$ the bottom 100$p$\% population has a weakly lower
percentage share of income under $F_a$ than under $F_b$. Formally, for
any two income distributions $F_a$ and $F_b$, we say that $F_b$ Lorenz
dominates $F_a$ if the Lorenz curve $L_{F_b}(p)$ associated with the
income distribution $F_b$ lies nowhere below that of Lorenz curve
$L_{F_a}(p)$ associated with the income distribution $F_a$ and at some
places (at least) lies above. Thus, we can think of domination
relation across pairs of Lorenz curves to infer about inequality and,
in particular, in a pairwise Lorenz curve comparison, higher of the
Lorenz curves are preferable. However, if the Lorenz curves of the two
distributions cross, then such an unambiguous conclusion about
inequality ordering cannot be drawn. The next example provides such an
instance of intersecting Lorenz curves.

\begin{ex}\label{ex1}
   Consider a society with four people and consider the following
  income distribution. Person 1 and Person 2 has an income of 20,
  Person 3 has an income of 30 and Person 4 has an income of 50. We
  first try to think of a meaningful representation of such an income
  distribution. Observe that if we draw a person at random, then with
  1/2 probability we will draw a person having an income of 20, with
  1/4 probability we will draw a person having an income of 30 and
  with 1/4 probability we will draw a person having an income of
  50. Therefore, we have a probability mass function of a random
  variable of three possible incomes $X_A=\{20,30,50\}$ and the
  probability mass function is given by $f_A(20)=1/2$, $f_A(30)=1/4$
  and $f_A(50)=1/4$. Using (\ref{examplezeroLorenz}), the Lorenz
  function is given by
	
	\begin{equation*}
	L_{F_A}(p) = \left\{\begin{array}{cl} \frac{2p}{3} & \mbox{ if
		$p\in \left(0,\frac{1}{2}\right]$,} \\ 
	\frac{6p-1}{6} & \mbox{if $p\in \left(\frac{1}{2},\frac{3}{4}\right]$,}\\
	\frac{5p-2}{3} & \mbox{if $p\in \left(\frac{3}{4},1\right]$.}
	\end{array} \right.
	\end{equation*}	
	
         Similarly, consider a society with four people and consider
        the following income distribution. Person 1 and Person 2 has
        an income of 15, Person 3 has an income of 42 and Person 4 has
        an income of 48. We have a probability mass function of a
        random variable $X_B=\{15,42,50\}$ and the probability mass
        function is given by $f_B(15)=1/2$, $f_B(42)=1/4$ and
        $f_B(48)=1/4$. Again, using (\ref{examplezeroLorenz}), the
        Lorenz function is given by
	
	\begin{equation*}
          L_{F_B}(p) = \left\{\begin{array}{cl} \frac{p}{2} & \mbox{ if
                                                              $p\in \left(0,\frac{1}{2}\right]$,} \\ 
                                \frac{28p-9}{20} & \mbox{if $p\in \left(\frac{1}{2},\frac{3}{4}\right]$,}\\
                                \frac{8p-3}{5} & \mbox{if $p\in \left(\frac{3}{4},1\right]$.}
	\end{array} \right.
    \end{equation*}

    Now consider the income distribution $F_A$ and compare it with the
    income distribution $F_B$. Note that at $p=1/2$,
    $L_{F_A}(1/2)=1/3>L_{F_B}(1/2)=1/4$ and at $p=3/4$,
    $L_{F_A}(3/4)=7/12<L_{F_B}(3/4)=3/5$. Hence, given both
    $L_{F_A}(p)$ and $L_{F_B}(p)$ are continuous in $p\in [0,1]$, the
    two Lorenz curves overlap and, in particular, these two Lorenz
    curve intersects at $p^*=17/24$, that is, at $p^*$ we have
    $L_{F_A}(p^*)=L_{F_B}(p^*)$.
  \end{ex}
  
  % An ordering based on the well-known Lorenz curve can be used for
  % checking whether different inequality indices can rank alternative
  % distributions of income in the same way.

  \section{Inequality indices in detail}
  
  \subsection{The Kolkata index} The $k$-index for any income
distribution $F$ is defined by the solution to the equation
$k_F + L_F(k_F) = 1$.  It has been proposed as a measure of income
inequality (see \cite{BCMM2020}, \cite{CG2017}, and, \cite{GC2014} for
more details). We can rewrite $k_F + L_F(k_F) = 1$ as
$\hat{L}_F(k_F)=k_F$ implying that the $k$-index is a fixed point of
the complementary Lorenz function. Since the complementary Lorenz
function maps $[0, 1]$ to $[0, 1]$ and is continuous, it has a fixed
point.  Furthermore, since complementary Lorenz function
$\hat{L}_F(p)$ is non-increasing, the fixed point is unique. Since for
any $F$, $p_F^*:=L_F^{-1}(1/2)\geq 1/2$ with the equality holding only
if we have an egalitarian income distribution, the unique fixed point
of $\hat{L}_F$ lies in the interval $[1/2, p_F^*]$ implying that for
any distribution $F$, $k_F\in [1/2,1)$. Therefore, $k_F$ lies between
50\% population proportion and the population proportion
$p^*_F=L_F^{-1}(1/2)$ that we associate with 50\% income given the
income distribution $F$. Observe that if $L_F(p) = p$, then
$k_F = 1/2$ and for any other income distribution, $1/2 < k_F <
1$. Also note that while the Lorenz curve typically has only two
trivial fixed points, that is, $L_F(0)=0$ and $L_F(1)=1$, the
complementary Lorenz function $\hat{L}_F(p)$ has a unique non-trivial
fixed point $k_F$.

The Pareto principle is based on Pareto's observation (in the year
1906) that approximately $80$\% of the land in Italy was owned by
$20$\% of the population.  The evidence, though, suggests that the
income distribution of many countries fails to satisfy the 80/20 rule
(see \cite{GC2014}).  The $k$-index can be thought of as a
generalization of the Pareto principle. Note that
$L_F(k_F) = 1 - k_F$; hence, the top $100(1 - k_F)$\% of the
population has $100(1 - (1 - k_F)) = 100k_F$\% of the income. Hence,
the ``Pareto ratio'' for the $k$-index is $k_F/(1 -k_F)$. Observe,
however, that this ratio is obtained endogenously from the income
distribution and in general, there is no reason to expect that this
ratio will coincide with the Pareto principle. The fact that the
$k$-index generalizes Pareto's 80/20 rule was first pointed out in
\cite{GC2014} and later also in \cite{EL2015A} and \cite{EL2016}.

\begin{itemize}
	\item \emph{Uniform distribution.} If we have the uniform distribution $F_u$ defined on $[a, b]$ where
	$0 \leq a < b < \infty$.  Then
	$$k_{F_u} = \frac{-(3a + b) + \sqrt{5a^2 + 6ab + 5b^2}}{2(b - a)}, \mathcal{K}_{F_u}=\frac{-2(a + b) + \sqrt{5a^2 + 6ab + 5b^2}}{(b - a)}.$$
      \item \emph{Exponential distribution.} For the exponential
        distribution $F_E$, the complementary Lorenz function is given
        by   $\hat{L}_{F_E}(p) = \\ (1 - p)\left[1 + \ln\left\{1/(1 -
            p)\right\}\right]$. One can show that
        $k_{F_E} \simeq 0.6822$ and hence
        $\mathcal{K}_{F_E}\simeq 0.3644$.
	\item \emph{Pareto distribution.} For the Pareto distribution
          $F_{P,\alpha}$, the complementary Lorenz function is given
          $\hat{L}_{F_{P,\alpha}}(p) = (1 - p)^{1 -
            \frac{1}{\alpha}}$.  The $k$-index is therefore a solution
          to (I) $(1 - k_{F_P})^{1 - \frac{1}{\alpha}} = k_{F_P}$. It
          is difficult to provide a general solution to (I). However,
          we an interesting observation in this context.

	\begin{itemize}
		   
        \item If $\hat{\alpha} = \ln 5/\ln 4\simeq 1.16$, then
          $k_{F_{P,\hat{\alpha}}}=0.8$ and we get the Pareto principle
          or the $80/20$ rule. Also note that
          $\mathcal{K}_{F_{P,\hat{\alpha}}}=0.6$
	\end{itemize} 
\end{itemize}

\subsubsection{Discrete random variable.} Consider any discrete random
variable with distribution function $F_G$ discussed in Example
\ref{drv} for which the Lorenz function is given by
(\ref{examplezeroLorenz}). To obtain the explicit form of the
$k$-index one can first apply a simple algorithm to identify the
interval of the form $[N(g-1),N(g))$ defined for $g\in \{1,\ldots,G\}$
in which the $k$-index can lie.

\noindent{\bf Algorithm-A:}
\begin{enumerate}
\item[Step 1:] Consider the smallest $g_1\in \{1,\ldots,G\}$ such that $N(g_1)\geq 1/2$ and consider the sum $N(g_1)+M(g_1)$. If $N(g_1)+M(g_1)\geq 1$, then stop and $k_{F_G}\in (N_{g_1-1},N(g_1)]$ and, in particular, $k_F=N(g_1)$ if and only if $N(g_1)+M(g_1)=1$. Instead, if $N(g_1)+M(g_1)<1$, then go to Step 2 and consider the group $g_1+1$ and repeat the process. \\
  $\vdots$
	
      \item[Step $t$:] We have reached Step $t$ means that in Step
        $(t-1)$ we had $N(g_1+t-1)+M(g_1+t-1)<1$. Therefore, consider
        the sum $N(g_1+t)+M(g_1+t)$. If $N(g_1+t)+M(g_1+t)\geq 1$, the
        stop and $k_{F_G}\in [N(g_1+t-1),N(g_1+t))$ and, in
        particular, $k_F=N(g_1+t)$ if and only if
        $N(g_1+t)+M(g_1+t)=1$. If $N(g_1+t)+M(g_1+t)<1$, then go to
        Step $(t+1)$.
        \end{enumerate}

         Observe that since $N(G)=M(G)=1$, if we have $N(G-1)+M(G-1)<1$
        in some step, then, in the next step, this algorithm has to
        end since $N(G)+M(G)=2>1$.

         Suppose for any discrete random variable with distribution
        function $F_G$ discussed in Example \ref{drv}, Algorithm-A
        identifies $g^*\in \{1,\ldots,G\}$ such that
        $N(g^*)+M(g^*)\geq 1$. If $N(g^*)+M(g^*)=1$, then
        $k_{F_G}=N(g^*)$ and if $N(g^*)+M(g^*)>1$, the $k_{F_G}$ is
        the solution to the following equation:
        
\begin{equation*}
  k_{F_G}+\left\{M(g^*-1)+\left(k_{F_G}-N(g^*-1)\right)\left(\frac{Nx_{g^*}}{M}\right)\right\}=1.
\end{equation*}

 Thus, to derive the $k$-index of any discrete random variable with
distribution function $F_G$ discussed in Example \ref{drv}, we first
identifying the group $g^*\in \{1,\ldots,G\}$ such that
$k_{F_G}\in (N(g^*-1),N(g^*)]$ (using Algorithm-A) and then, using
$g^*$, we get the following value of $k_{F_G}$:

\begin{equation*}
k_{F_G}= \left\{\begin{array}{cl} N(g^*) & \mbox{ if $N(g^*)+M(g^*)=1$,}\\
	
                  \frac{\mu_G+N(g^*)x_{g^*}-M(g^*)}{\mu_G+x_{g^*}} & \mbox{if $N(g^*)+M(g^*)>1$.}
\end{array} \right.
\end{equation*}	

\begin{rem}\label{rem1ne}
   Consider the income distributions $F_A$ and $F_B$ defined in
  Example \ref{ex1}. Recall that the Lorenz functions $L_{F_A}(p)$ and
  $L_{F_B}(p)$ are such that $L_{F_A}(p)>L_{F_B}(p)$ for all
  $p\in (0,17/24)$ and $L_{F_A}(p)<L_{F_B}(p)$ for all
  $p\in (17/24,1)$. However, one can work out that the $k$-indices for
  these distributions. Specifically, note that for $F_A$, $N(1)=1/2$
  and $M(1)=1/3$ implying that $N(1)+M(1)=1/6<1$ and $N(2)=3/4$ and
  $M(1)=7/12$ implying that $N(2)+M(2)=4/3>$. Hence, by Algorithm-A,
  $k_{F_A}\in (1/2,3/4)$ and it is a solution to the equation
  $k_{F_A}+(6k_{F_A}-1)/6=1$ implying that
  $k_{F_A}=7/12\simeq 0.58\dot{3}$ and hence the normalized value is
  $\mathcal{K}_{F_A}=1/6\simeq 0.1\dot{6}$.  Similarly, for $F_B$,
  $N(1)=1/2$ and $M(1)=1/4$ implying that $N(1)+M(1)=3/4<1$ and
  $N(2)=3/4$ and $M(1)=3/4$ implying that $N(2)+M(2)=3/2>$. Hence, by
  Algorithm-A, $k_{F_B}\in (1/2,3/4)$ and it is a solution to the
  equation $k_{F_B}+(28k_{F_B}-9)/20=1$ implying that
  $k_{F_B}=29/48\simeq 0.6041\dot{6}$ and hence the normalized value
  is $\mathcal{K}_{F_B}=5/24\simeq 0.208\dot{3}$.  Observe that
  $k_{F_A}<k_{F_B}$ and hence $\mathcal{K}_{F_A}<\mathcal{K}_{F_B}$
  implying that according to $k$-index as a measure of income
  inequality, the income distribution $F_A$ is less unequal than
  income distribution $F_B$.
\end{rem}

\subsubsection{The Hirsch index.} The physicist Jorge E.~Hirsch
suggested this index to measure the citation impact of the
publications of a research scientist (see \cite{Hirsch2005}). Let
$X = (x_1, \ldots ,x_m)$ be the set of research papers of a
scientist. Let $f:X \to \mathcal{N}$ be the \emph{citation function}
of the scientist.  The citation function simply gives the number of
citations for each publication.  Let $X_{()}=(x_{(1)},\ldots,x_{(m)})$
be a reordering of the elements in the set $X$ such that
$f(x_{(1)})\geq \ldots\geq f(x_{(m)})$. The Hirsch index, or the
$h$-index, is the largest number $H^*\in \{0, 1,\ldots,m\}$ such that
$f(x_{(H^*)})\geq H^*$. Note that if $f(x_{(1)})=0$, then $H^*=0$,
and, if $f(x_{(m)})\geq m$, then $H^*=m$ and for all other cases
$H^*\in \{1,\ldots,m-1\}$.

If neither $f(x_{(1)})=0$ nor $f(x_{(m)})\geq m$ holds, then how do we
identify the $h$-index? To see this, suppose that we plot a graph
where on the $x$-axis we plot the ordered set of publications of a
research scientist in non-increasing order of citations and on the
$y$-axis we plot the number of citations for each
publication. Moreover, if we join the consecutive plotted points like
$f(x_{(t)})$ and $f(x_{(t+1)})$ by a straight line for each
$t\in \{1,\ldots,m-1\}$, then we get a curve representing a function
$\tilde{f}:[1,m]\rightarrow [f(x_{1}),f(x_{m})]$, defined on the
domain $[1,m]$ with co-domain $[f(x_{1}),f(x_{m})]$, which we call the
{\it generated citation curve}. The generated citation curve is
continuous, piecewise linear and has a non-positive slope whenever the
slope exists. The generated citation curve resembles a lot like the
complementary Lorenz curve that we can associate with any income
distribution. Consider the fixed point of the generated citation curve
$\tilde{f}$ on the interval $[1,m]$, that is, consider
$\tilde{h}\in [1,m]$ such that $\tilde{f}(\tilde{h})=\tilde{h}$. As
long as there is at least one citation and as long as all papers are
not cited more than $(m-1)$-times, such a fixed point $\tilde{h}$
exists and is unique with the added property that
$\tilde{h}\in [1,m-1]$. Given this fixed point, we can identify the
relevant value of the $h$-index, that is, $H^*\in \{1,\ldots,m\}$ for
$f$ by the following procedure: If the fixed point $\tilde{h}$ is an
integer, then it is the $H^*$ that we are looking for, that is,
$\tilde{h}=H^*$. If, however, $\tilde{h}$ is not an integer, then
there exists an integer $\hat{h}$ such that
$\tilde{f}(x_{(\hat{h})})=f(x_{(\hat{h})})>\hat{h}$ and
$\tilde{f}(x_{(\hat{h}+1)})=f(x_{(\hat{h}+1)})<\hat{h}+1$ and then,
the relevant value of the $h$-index is $\hat{h}=H^*$. Therefore,
graphically, the procedure of obtaining the $h$-index of any research
scientist using the generated citation curve is the same as
identifying the fixed point of the complementary Lorenz function of
any income distribution that yields the $k$ index.

\subsection{The Gini index} The Gini index is the ratio of the area that lies between the
line of perfect equality and the Lorenz curve over the total area
under the line of perfect equality. If we plot cumulative share of
population from lowest income to highest income on the horizontal axis
and cumulative share of income on the Vertical axis (as shown in
Figure \ref{fig:Lorenz curve} above), then the Gini index
$\mathcal{G}_F(p)$ of any income distribution $F$ is given by
$\mathcal{G}_F:=\frac{area \ of \ AOCPA}{area \ of \ AOCBA}$. If all
people have non-negative income (or wealth, as the case may be), the
Gini index can theoretically range from 0 (complete equality) to
1 (complete inequality); it is sometimes expressed as a percentage
ranging between 0 and 100. In practice, both extreme values are not
quite reached. The Gini index is given by the following formula:
\begin{equation}\label{GCoefficient}
  \mathcal{G}_F=\frac{\int\limits_{0}^1(q-L_F(q))dq}{\left(\frac{1}{2}\right)}=2\int\limits_{0}^1(q-L_F(q))dq=1-2\int\limits_{0}^1L_F(q)dq.
\end{equation}

It is obvious that if $L_{F_e}(p)=p$ for all $p\in (0,1)$, then
$\mathcal{G}_F=0$. If the income distribution for a society with $n$
people follows a Power Law distribution, then $L_{F_n}(p)=p^n$. The
Gini index is then given by
$\mathcal{G}_{F_n}=\{1-2/(n+1)\}$. Hence, as $n\rightarrow \infty$, we
have $\mathcal{G}_{F_{\infty}}=1$. Gini index of some standard
continuous random variable are provided below.

\begin{itemize}
\item \emph{Uniform distribution:} Consider uniform distribution on
  some compact interval $[a,b]$ with $0\leq a<b<\infty$. The Gini
  index is given by
	\begin{equation*}              
          \mathcal{G}_{F_u}=2\int\limits_{0}^{1}\left[q-q\left\{1-\frac{(b-a)}{(a+b)}(1-q)\right\}\right]dq=\frac{(b-a)}{3(a+b)}>\mathcal{K}_{F_u}.
      \end{equation*}
      
    \item \emph{Exponential distribution:} Consider the exponential
      distribution with distribution function given by
      $F_E(x)=1-e^{-\lambda x}$ for any $x\geq 0$ with
      $\lambda>0$. The Gini index is given by
        
	\begin{equation*}
          \mathcal{G}_{F_E}=2\int\limits_0^1\left[q-L_{F_E}(q)\right]dq=2\int\limits_0^1(1-q)\ln\left(\frac{1}{1-q}\right)dq=\frac{1}{2}>\mathcal{K}_{F_E}. 
	\end{equation*}

      \item \emph{Pareto distribution:} For Pareto distribution given
        by the distribution function is
        $F_{P,\alpha}(x)=1-(m/x)^\alpha$ with $m>0$ as the minimum
        income and $\alpha>1$, the Gini index is given by
	\begin{equation*}
	\mathcal{G}_{F_{P,\alpha}}=2\int\limits_0^1\left[q-\left\{1-(1-q)^{1-\frac{1}{\alpha}}\right\}\right]dq=\frac{1}{2\alpha-1}. 
      \end{equation*}

      If we plot the Gini index for different values of $\alpha>1$,
      then note that as $\alpha$ increases the Gini index
      decreases, and, as $\alpha\rightarrow 1$ we have
      $\mathcal{G}_{F_{P,\alpha}}\rightarrow 1$. Also note that if
      $\hat{\alpha} = \ln 5/\ln 4$, then
      $\mathcal{G}_{F_{P,\hat{\alpha}}}\simeq
      0.7565>\mathcal{K}_{F_{P,\hat{\alpha}}}=0.6$.
\end{itemize}

\subsubsection{Discrete random variable.} Consider the discrete random variable $F_G$ discussed in Example
\ref{drv} for which the Lorenz function is given by
(\ref{examplezeroLorenz}). As show in Appendix A, we have the
following explicit form of the Gini index.

\begin{equation}\label{discreteexpression}
\mathcal{G}_{F_G}=\frac{\sum\limits_{g=1}^G\sum\limits_{t=1}^{G}n_tn_g|x_t-x_g|}{2NM}.
\end{equation}

Note that if $n_g=1$ for all $g\in \{1,\ldots,G\}$ so that $G=N$ and
$M=\sum_{g=1}^Nx_g$, then from (\ref{discreteexpression}) it follows
that

\begin{equation}\label{discreteexpressionspecial}
  \mathcal{G}_{F_N}=\frac{\sum\limits_{g=1}^N\sum\limits_{t=1}^{N}|x_t-x_g|}{2N\sum\limits_{g=1}^Nx_g}.
\end{equation} 

\begin{rem}\label{rem2wo}
   Consider the income distributions $F_A$ and $F_B$ defined in
  Example \ref{ex1}. One can work out that the Gini indices are
  $\mathcal{G}_{F_A}=\mathcal{K}_{F_B}=5/24\simeq
  0.208\dot{3}>\mathcal{K}_{F_A}$ and
  $\mathcal{G}_{F_B}=21/80=0.2625>\mathcal{K}_{F_B}$. Hence, like the
  normalized $k$-index, according Gini index the income
  distribution $F_A$ is less unequal than income distribution $F_B$.
\end{rem}

\subsection{The Pietra index} An interesting index of inequality is the Pietra index (see Pietra
\cite{Pi1915}) that tries to identify that proportion of total income
that needs to be reallocated across the population in order to achieve
perfect equality. Given any income distribution $F$, this proportion
is given by the maximum value of $p-L_F(p)$. Therefore, the Pietra
index is $\mathcal{P}_F=\max_{p\in [0,1]}(p-L_F(p))$. It is immediate
that if $L_F(p)=p$ for all $p\in [0,1]$, then
$\mathcal{K}_F=\mathcal{P}_F=\mathcal{G}_F=0$. For any other income
distribution $F$, the maximum distance between the perfect equality
line and the Lorenz curve is the distance OP in Figure \ref{fig:Lorenz
  curve} above. Note that for any random variable $X$ with
distribution function $F$,
$p-L_F(p)=p-\left(\int_0^pF^{-1}(q)dq\right)/\mu=\int_0^p\left\{\mu-F^{-1}(q)dq\right\}/\mu$. Therefore,
maximizing $(p-L_F(p))$ by selecting $p\in [0,1]$ is equivalent to
maximizing the area $\int_{0}^p\left\{\mu-F^{-1}(q)\right\}dq$ by
selecting $p\in [0,1]$. Since the Lorenz curve plots the percentage of
total income earned by various portions of the population when {\it
  the population is ordered by the size of their incomes}, it is
obvious that $\left\{\mu-F^{-1}(q)\right\}>0$ for all
$q\in [0,F(\mu))$, $\left\{\mu-F^{-1}(q)\right\}<0$ for all
$q\in (F(\mu),1]$ and $\left\{\mu-F^{-1}(q)\right\}=0$ at
$q=F(\mu)$. Thus, it follows that the maximum value of the integral
$\int_{0}^p\left\{\mu-F^{-1}(q)\right\}dq$ is attained at
$p=F(\mu)$. Hence, the Pietra index for any random variable with
distribution function $F$ is
\begin{equation}
\mathcal{P}_F=\max_{p\in [0,1]}(p-L_F(p))=F(\mu)-L_F(F(\mu)). 
\end{equation} 

\begin{itemize}
\item \emph{Uniform distribution:} For the uniform distribution on
  some compact interval $[a,b]$ with $0\leq a<b<\infty$, we have
  $p-L_{F_u}(p)=(b-a)p(1-p)/(a+b)$ for all $p\in [0,p]$. Moreover,
  $\mu_u=(a+b)/2$ and as a result $F_u(\mu_{u})=1/2$. Hence, the
  Pietra index is given by
  
	\begin{equation*}              
          \mathcal{P}_{F_u}=\frac{(b-a)}{(a+b)}F_u(\mu_u)(1-F_u(\mu_u))=\frac{(b-a)}{4(a+b)}.
      \end{equation*}

      Given $\mathcal{G}_{F_u}=(b-a)/3(a+b)$, we have
      $\mathcal{P}_{F_u}=(3/4)\mathcal{G}_{F_u}<\mathcal{G}_{F_u}$. Moreover,
      one can easily check that $\mathcal{P}_{F_u}>\mathcal{K}_{F_u}$.
      
      \item \emph{Exponential distribution:} For the the exponential
        distribution $F_E(x)=1-e^{-\lambda x}$ defined for any
        $x\geq 0$ with $\lambda>0$, we have
        $p-L_E(p)=(1-p)\ln (1/(1-p))$ for all $p\in [0,1]$. We also
        have $\mu_E=1/\lambda$ and hence $F_E(\mu_E)=1-e^{-1}$. The
        Pietra index is given by

          \begin{equation*}
	\mathcal{P}_{F_E}=(1-F_E(\mu_E))\ln \left(\frac{1}{1-F_E(\mu_E)}\right)=\frac{1}{e}. 
      \end{equation*}

      Observe that
      $\mathcal{K}_{F_E}\simeq0.3644<\mathcal{P}_{F_E}=1/e\simeq
      0.3679<\mathcal{G}_{F_E}=1/2$.
      
      \item \emph{Pareto distribution:} For Pareto distribution given
        by the distribution function is
        $F_{P,\alpha}(x)=1-(m/x)^\alpha$ with $m>0$ as the minimum
        income and $\alpha>1$, we have
        $p-L_P(p)=(1-p)^{1-\frac{1}{\alpha}}-(1-p)$ for all
        $p\in [0,p]$, $\mu_P=\alpha m/(\alpha-1)$ and
        $F_{P,\alpha}(\mu_P)=1-\{(\alpha-1)/\alpha\}^\alpha$. The
        Pietra index is given by

          \begin{equation*}
            \mathcal{P}_{F_{P,\alpha}}=(1-F_P(\mu_P))^{1-\frac{1}{\alpha}}-(1-F_P(\mu_P))=\frac{(\alpha-1)^{\alpha-1}}{\alpha^\alpha}. 
      \end{equation*}

      One can verify that
      $\mathcal{P}_{F_{P,\alpha}}<\mathcal{G}_{F_P}=1/(2\alpha-1)$ for
      all $\alpha>1$. Also note that if $\hat{\alpha} = \ln 5/\ln 4$,
      then
      $\mathcal{G}_{F_{P,\hat{\alpha}}}\simeq
      0.7565>\mathcal{P}_{F_{P,\hat{\alpha}}}\simeq
      0.626655>\mathcal{K}_{F_{P,\hat{\alpha}}}=0.6$.
\end{itemize}

 As shown in Appendix B (i), there is an alternative representation of
the Pietra index as the ratio of the mean absolute deviation of the
income distribution and twice its mean, that is,
$\mathcal{P}_F=E(|x-\mu|)/2\mu$.

\subsubsection{Discrete random variable.} Consider the discrete random
variable $F_G$ discussed in Example \ref{drv} for which the Lorenz
function is given by (\ref{examplezeroLorenz}). It is shown in
Appendix B (ii) that the Pietra index has the following
representations:

\begin{equation}\label{Pietra2forms}
  \mathcal{P}_{F_G}=\frac{\sum\limits_{g=1}^{\tilde{g}}n_g\left(\mu_G-x_g\right)}{M}=\frac{E(|x-\mu_G|)}{2\mu_G},  
\end{equation}

where $\tilde{g}\in \{1,\ldots,G-1\}$ is such that
$\mu_G\in [x_{\tilde{g}},x_{\tilde{g}+1})$ implying that
$F_G(\mu_G)=N(\tilde{g})$.

\begin{rem}\label{rem3ee}
   Consider the income distributions $F_A$ and $F_B$ defined in
  Example \ref{ex1}. Observe that for both $F_A$ and $F_B$ the mean is
  the same and, in particular $\mu_A=\mu_B=30$. Therefore,
  $F_A(\mu_A)=3/4$ and $L_{F_A}(\mu_A)=7/12$ implying
  $\mathcal{P}_{F_A}=\mathcal{K}_{F_A}=1/6\simeq
  0.1\dot{6}<\mathcal{G}_{F_A}$, and, we also have $F_B(\mu_B)=1/2$
  and $L_{F_B}(\mu_A)=1/4$ implying
  $\mathcal{P}_{F_B}=1/4=0.25\in
  (\mathcal{K}_{F_B},\mathcal{G}_{F_B})$. Thus,
  $\mathcal{P}_{F_A}<\mathcal{P}_{F_B}$ and hence, like the ordering
  with the $k$-index as well as the Gini index, according to
  the Pietra index, the income distribution $F_A$ is less unequal than
  income distribution $F_B$.
\end{rem}

\section{Comparing the measures}

\subsection{Rich-poor disparity}

The Gini index, as is well-known, measures inequality by the
area between the Lorenz curve and the line of perfect equality.  For
any $p\in [0,1]$, one can decompose the Gini index into three
parts: two representing the \emph{within-group inequality} and one
representing the \emph{across-group inequality}.  In Figure
\ref{fig:disparity} below, the unshaded area bounded by the Lorenz
curve and the line from $(0, 0)$ to $(p, L_F(p))$ is the within-group
inequality of the poor. It represents the extent to which inequality
can be reduced by redistributing incomes among the poor.  Similarly,
the area bounded by the Lorenz curve and the line segment from
$(p, L_F(p))$ to $(1, 1)$ represents the within-group inequality of
the rich.  The shaded area represents the across-group
inequality. Easy computation shows that the extent of across-group
inequality between the bottom $p\times100\%$ and top
$(1-p)\times 100\%$ is the (across-group) disparity function
$D_F(p) = (1/2)[p - L_F(p)]$.  One can ask for what value of $p$ is
the across-group inequality maximized?  The answer is that this is
maximized at the proportion associated with the \emph{Pietra index}
given by $\mathcal{P}_F=F(\mu)-L_F(F(\mu))$. Hence, $F(\mu)$ is the
proportion where the disparity is maximized. Therefore, the Pietra
index is that fraction which splits the society into two groups in a
way such that inter-group inequality is maximized.

\begin{figure}[h]
	\begin{center}
		\begin{tikzpicture}
		\begin{axis}[xlabel={fraction of population}, 
		ylabel={fraction of income},
		xmin=0,
		ymin=0,
		xmax=1,
		ymax=1,
		legend style={
			at={(0.25,0.95)},
			anchor=north,
		}]
		\addplot[name path=A,color=red, thick, domain=0:1] {x*x};
		\legend{Lorenz curve};
		\addplot[domain=0:1] {x};
		\addplot coordinates
		{(0, 0) (0.5, 0.25) (1, 1)} --cycle [fill=gray];
		\node[right] at (0.51, 0.25){\small $(P, L(P))$};
		\addplot[name path=B,color=black,domain=0:0.5]{x};
		\addplot[name path=C,color=red, thick, domain=0:0.5] {x*x};
		\addplot[blue!50] fill between[of=C and B];
		\addplot[name path=B,color=black,domain=0.5:1]{x};
		\addplot[name path=C,color=red, thick, domain=0.5:1] {x*x};
		\addplot[green!50] fill between[of=C and B];
		\end{axis}
		\end{tikzpicture}
		\caption{Rich-poor disparity
                  assuming that the poor are $p$\% of the population.
                  The blue-shaded area is the disparity among the
                  poor, the green-shaded area is the disparity among
                  the rich, and the grey-shaded area is the disparity
                  between the rich and the poor.}
		\label{fig:disparity}
	\end{center}
\end{figure}
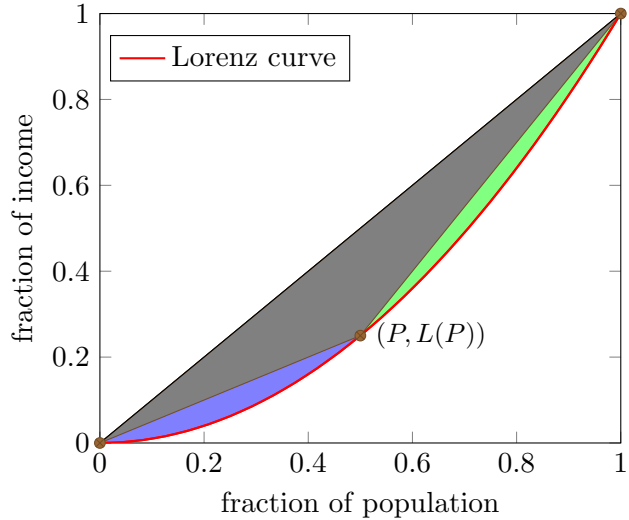

The discussion to follow shows that interpretation of the $k$-index is
different from that of the Pietra index. Let us divide society into
two groups, the ``poorest'' who constitute a fraction $p$ of the
population and the ``richest'' who constitute a fraction $1 - p$ of
the population.  Given the Lorenz curve $L_F(p)$, we look at the
distance of the ``boundary person'' from the poorest person on the one
hand and the distance of this person from the richest person on the
other hand.  These distances are given by $\sqrt{p^2 + L_F(p)^2}$ and
$\sqrt{(1 - p)^2 + (1 - L_F(p))^2}$ respectively.  Then, the $k$-index
divides society into two groups in a manner such that the Euclidean
distance of the boundary person from the poorest person is equal to
the distance from the richest person.

 The value of the disparity function at the $k$-index is
$D_F(k_F) = k_F - 1/2$. It measures the gap between the proportion
$k_F$ of the poor from the $50-50$ population split.  As long as we do
not have a completely egalitarian society, $k_F > 1/2$ and hence it is
one way of highlighting the rich-poor disparity with $k_F$ defining
the income proportion of the top $(1-k_F)$ proportion of the rich
population. In terms of disparity, the Gini index and Pietra
index do not have as nice an interpretation.

\subsection{Comparison of magnitudes} To compare the $k$-index with other measures of
inequality we will use the normalized $k$-index which is given by
$\mathcal{K}_F:=2k_F-1$. The normalized $k$-index was first introduced
in \cite{EL2015A} and was called the ``perpendicular-diameter index''
(see \cite{EL2015A}, \cite{EL2015B}, \cite{EL2016}). For all income
distributions used till the previous section we found that given any
$F$, the value of the normalized $k$ index is no more than the value
of the Pietra index and the value of the Pietra index is no more than
the value of the Gini index. This is not just a coincidence. It
was established in \cite{BCMM2020} that for any income distribution
$F$, we have $\mathcal{K}_{F}\leq \mathcal{P}_F\leq \mathcal{G}_F$. It
is obvious that since the Pietra index maximizes $p-L_F(p)$, it is
obvious that $\mathcal{K}_{F}=2k_F-1=k_F-L_F(k_F)\leq \mathcal{P}_F$.
Moreover, in \cite{BCMM2020}, it was also established that for any
given distribution $F$ and any $p\in [0,1]$,
$p-L_F(p)\leq \mathcal{G}_F$ and hence, using this result, it follows
that $\max_{p\in [0,1]}\{p-L_F(p)\}\leq \mathcal{G}_F$ and hence we
get $\mathcal{P}_F\leq \mathcal{G}_F$.

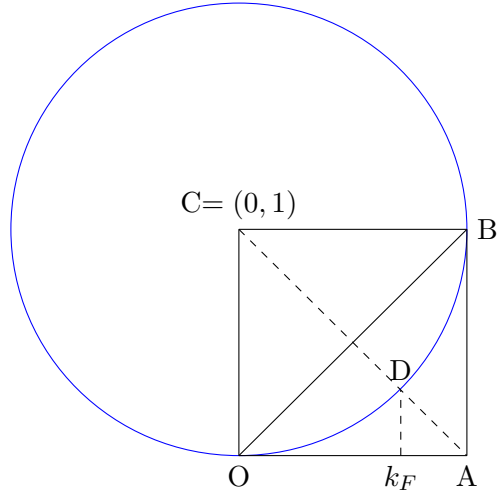
\begin{figure}[h]
	\begin{center}
		\begin{tikzpicture}[scale=3]
		
		\draw [blue](1,1)circle(1cm);
		\draw (1,1)--(1,0) node [pos=0,above ] {C$=(0,1)$}  node [pos=1,below ] {O};
		\draw (1,1)--(2,1)   node [pos=1, right] {B};
		\draw (2,0)--(2,1) node [pos=0,below ] {A} ;
		\draw (1,0)--(2,0);
		\draw (1,0)--(2,1);
		\draw [dashed](1,1)--(2,0);
		\draw [dashed](1.71,0)--(1.71,0.29) node [pos=0,below] {$k_F$}  node [pos=1,above] {D};
		\end{tikzpicture}
		\caption{The Lorenz curve as an arc
                    of a unit circle. Here, the normalised $k$-index
                    and Pietra index are equal but different from the
                    Gini index:
                    $\mathcal{K}_{F_R}= \mathcal{P}_{F_R}
                    =\sqrt{2}-1<G=\pi/2-1$.}
		\label{circle}
	\end{center}
      \end{figure}

We first provide an example where the normalized $k$-index coincides
with the Pietra index. This example is taken from \cite{BCMM2020}. Let
us consider an arc of a unit circle ODB as a Lorenz curve where OB is
one of the diagonal(egalitarian line) of the unit square ABCO (as
shown in Figure \ref{circle}) where CD represents the unit radius of
the circle, CA is the other diagonal of the unit square ABCO =
$\sqrt{2}$. In this case the Lorenz curve is,
$L_{F_{kg}}(p)= 1-\sqrt{1-p^2}$ where $F_{kg}$ is the relevant income
distribution. One can verify that
$\mathcal{K}_{F_{kg}}=\mathcal{P}_{F_{kg}}=\sqrt{2}-1\simeq
0.4142<\mathcal{G}_{F_{kg}}=(\pi/2)-1\simeq 0.571$. Hence, the Gini
insdex is larger than the Pietra index and the normalized
$k$-index. Also in this case the maximim distance between perfect
equality line and the Lorenz curve is at
$k_{F_{kg}}=F(\mu_{kg})=1/\sqrt{2}$, hence Pietra index coincides with
the normalized $k$-index.

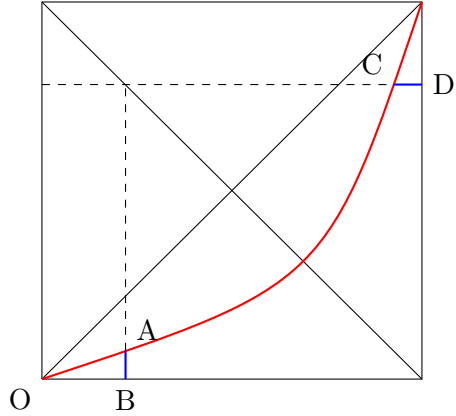
\begin{figure}[h]
	\begin{center}
		\begin{tikzpicture}[scale=5]
		\draw (0,0)--(0,1)--(1,1)--(1,0)--(0,0);
		\draw (0,0)--(1,1) node [pos=0,below left] {O}  ;
		\draw (0,1)--(1,0) ;
		\draw [red,thick](0,0)..controls(0.75,0.25)..(1,1);
		\draw [blue,thick](0.22,0)--(0.22,0.075)node[pos=0,below,black]{B} node[pos=1,above right,black]{A};
		\draw [dashed](0.22,0.075)--(0.22,0.781);
		\draw [dashed](0,0.781)--(0.925,0.781);
		\draw [blue,thick](0.925,0.781)--(1,0.781)node[pos=0,above left,black]{C} node[pos=1, right,black]{D};
		\end{tikzpicture}
		\caption
                {Lorenz curve for which Pietra
                    index and normalised $k$-index are equal. The
                    similarity holds only when for all $p \in [0, 1]$,
                    $\overline{AB} = \overline{CD}$, where
                    $A\equiv(p,L_F(p)), B \equiv (p, 0),
                    C\equiv(L^{-1}_F(1 - p), (1-p))$ and
                    $D\equiv(1, 1 - p)$.}
		\label{symmetryLorenzF}
	\end{center}
\end{figure}

      The Lorenz function $L_F(p)$ is \emph{symmetric} if for all
      $p\in [0,1]$, $L_F(\hat{L}_F(p))=1-p$ or equivalently
      $L_F(p)+r_F(p)=1$, where $r_F(p)=L^{-1}_F(1-p)$. The idea of
      symmetry is explained in Figure \ref{symmetryLorenzF}. One can
      verify that the Lorenz function $L_{F_{kg}}(p)= 1-\sqrt{1-p^2}$
      is symmetric. It was proved in Banerjee, Chakrabarti, Mitra and
      Mutuswami \cite{BCMM2020} that, in general, if the Lorenz
      function is symmetric and differentiable, then the proportion
      $F(\mu)$ associated with the Pietra index coincides with the
      proportion $k_F$ of the $k$-index. Hence, we also have
      $\mathcal{K}_F=\mathcal{P}_F$.

The next example is one where the Pietra index coincides with the Gini
index. This example is taken from Eliazar and Sokolov
\cite{ES2010}. Fix any fraction $x_0\in (0,1)$ and consider the
following Lorenz function:
\begin{equation}
L_{F_{pg}}(p)=
\left\{ \begin{array}{ll}
0  & \mbox{if $p\in [0,x_0]$,} \\
\frac{(p-x_0)}{(1-x_0)}  & \mbox{if $p\in (x_0,1]$.} 
\end{array}
\right.  
\end{equation}

\begin{figure}[h]
	\begin{center}
		\begin{tikzpicture}[scale=5]
                  \draw (0,0)--(0,1)--(1,1)--(1,0)--(0,0);
                  \draw[white](0,0)--(0.7,0); \draw (0,0)--(1,1)
                  node[pos=0,below left] {O} node [pos=1,above right]
                  {A}; \draw (0,1)--(1,0) node [pos=1,below right]
                  {C}; \draw (0.7,0)--(1,1) node[pos=0,below, black]
                  {$a$} node[pos=0,above left,thick,black]{B}; \draw
                  [thick,red](0,0)--(0.7,0)--(1,1);
                  \draw[dashed](0.77,0)--(0.77,0.23) node[pos=0,below
                  right]{$k$} node[pos=1,above,right]{Q};
		\end{tikzpicture}
	\end{center}
	\caption{A Lorenz curve depicting two
            groups, one with no income and the other where all agents
            have the same income. The Gini index and the Pietra
            index are equal but different from the normalised
            $k$-index: $G = P = x_0 >\mathcal{K}$.} \label{GPC}
\end{figure}
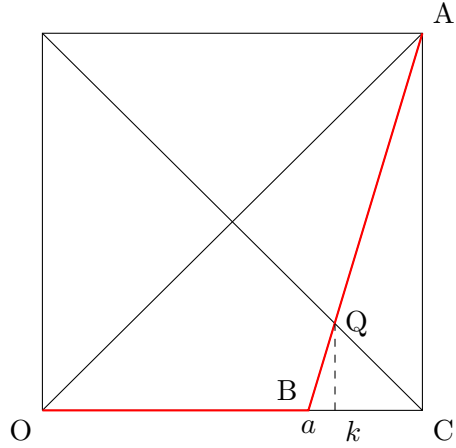

Figure \ref{GPC} depicts this Lorenz function $L_{F_{pg}}(.)$ and in
particular the curve OBA represents this Lorenz curve. One can show
that
$\frac{x_0}{2-x_0}=\mathcal{K}_{F_{pg}}<\mathcal{P}_{F_{pg}}=\mathcal{G}_{F_{pg}}=x_0$. Hence,
the Gini index coincides with Pietra and the normalized $k$-index has
a lower value. Therefore, from this example we can say that $k$-index
has different features relative to both the Gini index and the Pietra
index.

 Finally, when does all the three indices coincide? It was established
in \cite{BCMM2020} that all three measures will coincide if and only
if the Lorenz function has the following form defined for any given
$C\in [1/2,1)$:
\begin{equation}
L_C(p)=
\left\{ \begin{array}{ll}
\left(\frac{1-C}{C}\right)p  & \mbox{if $p\in [0,C]$,} \\
(1-C)+\frac{C}{(1-C)}(p-C)  & \mbox{if $p\in(C,1]$.} 
\end{array}
\right.  
\end{equation}

\begin{figure}[h]
	\begin{center}
		\begin{tikzpicture}[scale=5]
                  \draw (0,0)--(0,1)--(1,1)--(1,0)--(0,0);
                  \draw (0,0)--(1,1) node [pos=0,below left] {O} node
[pos=1,above right] {B};
                  \draw (0,1)--(1,0)node [pos=0,above left] {C} node
[pos=1,below right] {A};
		\draw [red,thick](0,0)--(0.7,0.3) node [pos=1,right,
                black] {Q};
                \draw [red,thick](0.7,0.3)--(1,1);
                \draw [dashed](0.7,0.3)--(0.7,0) node
[pos=1,below]{$k_C=C$};
		\end{tikzpicture}
	\end{center}
	\caption{A Lorenz curve depicting two
            groups with equally distributed incomes but differing
            average incomes. The Gini, Pietra and normalised $k$
            indices are all equal here:
            $\mathcal{K}_{F_C}=\mathcal{P}_{F_C}=\mathcal{G}_{F_C}=
            2k_F - 1$.} \label{LFCKGP}
\end{figure}
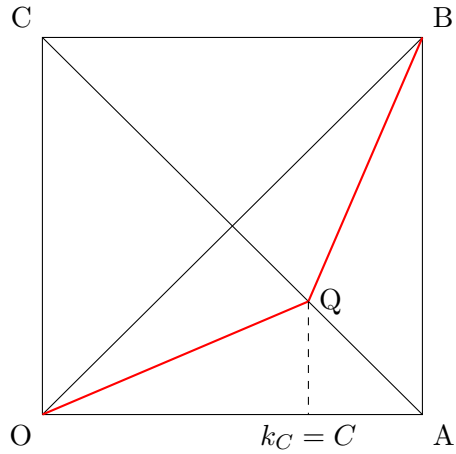

In Figure \ref{LFCKGP}, the straight lines OQ and QB taken together
represents the Lorenz curve for $L_C(.)$. One can verify that

\begin{equation}\label{eq:gkrel}
    \mathcal{K}_{F_C}=\mathcal{P}_{F_C}=\mathcal{G}_{F_C}=2C-1
\end{equation}

Observe that, if $C=1/2$, then we have $L_{F_{0.5}}(p)=L_{F_e}(p)=p$
for all $p\in (0,1)$ and in that case the three indices also coincide
since $\mathcal{G}_{F_e}=\mathcal{P}_{F_e}=\mathcal{K}_{F_e}=0$.

It is clear that the Lorenz functions of the form $L_{F_C}(.)$ with
$C\in (1/2,1)$ is valid for any society having two income
groups. Therefore, a natural question in this context is the
following: What does the coincidence of the three measures mean in
terms of discrete random variables? For any discrete random variable
$F_{G}$ such that $G=2$, we have $N=n_1+n_2$, $M=n_1x_1+n_2x_2$ with
$x_1<x_2$ and the associated Lorenz function has the following form:

\begin{equation}\label{dtv}
  L_{F_2}(p)=
  \left\{ \begin{array}{ll}
            \frac{(n_1+n_2)x_1p}{n_1x_1+n_2x_2}  & \mbox{if $p\in \left(0,\frac{n_1}{n_1+n_2}\right]$,} \\
            \frac{n_1x_1}{n_1x_1+n_2x_2}+\left(\frac{(n_1+n_2)x_2}{n_1x_1+n_2x_2}\right)\left(p-\frac{n_1}{n_1+n_2}\right)  & \mbox{if $p\in \left(\frac{n_1}{n_1+n_2},1\right)$.} 
\end{array}
\right.  
\end{equation}

 For the coincidence of all the three indices we first require that
$C\in (1/2,0)$ and $C=n_1/(n_1+n_2)$ implying that
$n_1>n_2$. Moreover, for the coincidence we also require $C=k_{F_2}$,
that is, $C+L_{F_{2}}(C)=1$ which yields $n^2_1x_1=n^2_2x_2$. Thus,
from the above discussion we have the following result.

\begin{itemize}
\item Consider any discrete random variable $F_G$ discussed in Example
  \ref{drv} for which the Lorenz function is given by
  (\ref{examplezeroLorenz}). The normalized $k$-index coincides with
  the Gini index and the Pietra index if and only if any one of
  the following conditions holds:
	\begin{enumerate}
        \item[(C1)] The society has all agents having the same income
          $x_1>0$ so that $L_{F_1}(p)=L_{F_e}(p)=p$ for all
          $p\in (0,1)$.  For this case we have,
          $\mathcal{K}_{F_1}=\mathcal{P}_{F_1}=\mathcal{G}_{F_1}=0$.
        \item[(C2)] The society has two groups of agents with one
          group of $n_1$ agents having an income of $x_1$ and another
          group of $n_2$ agents having an income of $x_2$ such that
          $x_1<x_2$. Moreover, the Lorenz function is $L_{F_2}(p)$
          given in (\ref{dtv}) with the added restrictions that
          $n_1>n_2$, $n_1^2x_1=n_2^2x_2$ and hence
          $n_1x_1<n_2x_2$. For this case we have,
          $\mathcal{K}_{F_2}=\mathcal{P}_{F_2}=\mathcal{G}_{F_2}=2k_{F_2}-1=(n_1-n_2)/(n_1+n_2)$.
	\end{enumerate}
\end{itemize}

\section{Ranking Lorenz functions}
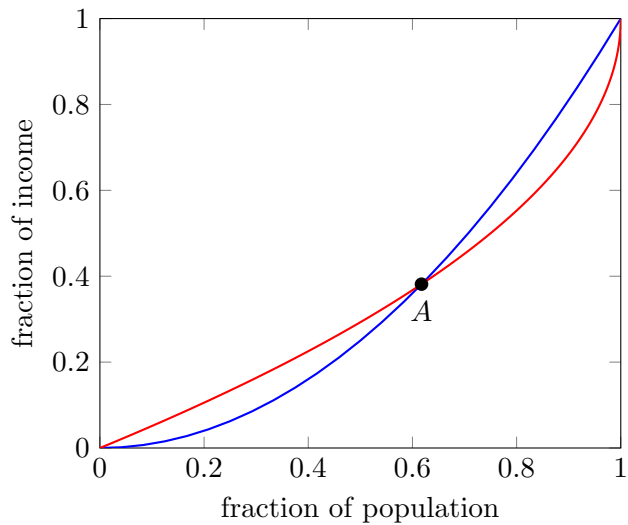
\begin{figure}[h]
	\begin{center}
		\begin{tikzpicture}
		\begin{axis}[xlabel={fraction of population}, 
		ylabel={fraction of income},
		xmin=0,
		ymin=0,
		xmax=1,
		ymax=1,
		legend style={
			at={(0.25,0.95)},
			anchor=north,
		}]
		\addplot[name path=A,color=blue, thick, domain=0:1] {x*x};
		\addplot[name path=B,color=red, thick, domain=0:1,samples=1000] {1-sqrt(1-x)};
		\fill [name intersections={of=A and B,by={E1,E2}}] (E2)
		node[circle,fill,inner sep=1.8pt,label=below:$A$]{};
		\end{axis}
		\end{tikzpicture}
		\caption{Two Lorenz curves with identical Gini, Pietra and normalized $k$-indices. The blue curve is $L_{F_{\bar{u}}}(p) = p^2$ and the red curve is $L_{F_{P,2}}(p) =1 - \sqrt{1 -
                        p}$.}
    	\label{fig:EquivFig}
	\end{center}
\end{figure}

 \noindent Consider the uniform income distribution $F_{\bar{u}}$ defined on any
compact interval $[0,b]$ with $b>0$. The Lorenz function is given by
$L_{F_{\bar{u}}}(p) = p^2$ for all $p\in [0,1]$ (see Figure
\ref{fig:EquivFig}). Here $k_{F_{\bar{u}}}$ is the reciprocal of the
Golden ratio, that is, $k_{F_{\bar{u}}} = (\sqrt{5}-1)/2 = 1/\phi$
where $\phi = (\sqrt{5}+1)/2\simeq 0.61803$ is the \emph{Golden
  ratio}. Moreover, $\mathcal{K}_{F_u}=\sqrt{5}-2\simeq
0.23607$. Similarly, one can derive that the Gini index is
$\mathcal{G}_{F_{\bar{u}}}=1/3$ and the Pietra index is
$\mathcal{P}_{F_{\bar{u}}}=1/4$ with $\mu_{\bar{u}}=1/2$. Hence, we
have
$\mathcal{G}_{F_{\bar{u}}}=1/3>\mathcal{P}_{F_{\bar{u}}}=1/4>\mathcal{K}_{F_{\bar{u}}}=\sqrt{5}-2$. Similarly,
consider the Pareto distribution $F_{P,\alpha}$ with parameter value
$\alpha=2$. The Lorenz function is given by
$L_{F_{P,2}}(p) =1-\sqrt{1 - p}$ so that
$\hat{L}_{F_{P,2}}(p) =\sqrt{1 - p}$ and the $k$-index is again the
reciprocal of the Golden ratio, that is, $k_{F_{P,2}}=1/\phi$ and
$\mathcal{K}_{F_{P,2}}=\sqrt{5}-2$ (see Figure
\ref{fig:EquivFig}). Thus, according to the normalized $k$-index, a
society with an income distribution $F_{\bar{u}}$ is equivalent to a
society with an income distribution of $F_{P,2}$ in terms of
inequality. One can verify that this equivalence between $F_{\bar{u}}$
and $F_{P,2}$ is also preserved under the Gini index and the
Pietra index. Specifically, we have
$\mathcal{G}_{F_{P,\alpha}}=\mathcal{G}_{F_{\bar{u}}}=1/3$ and
$\mathcal{P}_{F_{P,2}}=\mathcal{P}_{F_{\bar{u}}}=1/4$ though
$\mu_{P,2}=3/4>\mu_{\bar{u}}=1/2$. Hence, we
have
$$\mathcal{G}_{F_{P,\alpha}}=\mathcal{G}_{F_{\bar{u}}}=1/3>\mathcal{P}_{F_{P,2}}=\mathcal{P}_{F_{\bar{u}}}=1/4>\mathcal{K}_{F_{P,\alpha}}=\mathcal{K}_{F_{\bar{u}}}=\sqrt{5}-2.$$

%\begin{figure}[h]
	%\begin{center}
%\includegraphics[scale=0.6]{graph.png}
%\caption {Plot of $L_{F_{\bar{u}}}(p) = p^2$ in blue and $L_{F_{P,2}}(p) =1-\sqrt{1 - p}$ in red}\label{fig:EquivFig}
%\end{center}
%\end{figure}

 Consider the income distributions $F_A$ and $F_B$ defined in Example
\ref{ex1}. From Remark \ref{rem1ne} it follows that
$\mathcal{K}_{F_A}<\mathcal{K}_{F_B}$, from Remark \ref{rem2wo} it
follows that $\mathcal{G}_{F_A}<\mathcal{G}_{F_B}$ and from Remark
\ref{rem3ee} it also follows that
$\mathcal{P}_{F_A}<\mathcal{P}_{F_B}$. Therefore, all the three
measures unambiguously assures that the society with income
distribution $F_{A}$ is less unequal that the society with income
distribution $F_{B}$.

 Given the above examples of this section, one may be tempted to think
that ranking Lorenz functions using these three measures always gives
the same order, that is, if one measure shows that the income
distribution $F$ is equivalent to another income distribution $F'$ in
terms of inequality, then the other two measures will also give the
same result, and, if one measure shows that the income distribution
$F$ is less unequal than the income distribution $F'$, then also the
other two measures will establish the same order. However, as argued
in \cite{BCMM2020}, this is not the case. To establish this point
\cite{BCMM2020} provided the following two examples.

In the first example the following Lorenz functions were considered to
establish that the normalized $k$-index yields a different ranking
from the Pietra index.

	\begin{equation*}
	L_{F_a}(p)=
	\left\{ \begin{array}{ll}
	\frac{3p}{4}  & \mbox{if $p\in [0,1/3]$,} \\
	\frac{9p-1}{8}   & \mbox{if $p\in (1/3,1]$.} 
	\end{array}
	\right.  
	\end{equation*} 
	
	\begin{equation*}
	L_{F_b}(p)=
	\left\{ \begin{array}{ll}
	\frac{8p}{9}  & \mbox{if $p\in [0,7/8]$,} \\
	\frac{16p-7}{9}  & \mbox{if $p\in (7/8,1]$.} 
	\end{array}
	\right.  
      \end{equation*}

      One can show that
      $\mathcal{K}_{F_a} = \mathcal{K}_{F_b} = 1/7 < \mathcal{P}_{F_a}
      = 1/12 <\mathcal{P}_{F_b} = 7/72$, that is, according to the
      normalized $k$-index, the society with income distribution $F_a$
      is equivalent to the society with income distribution $F_b$ in
      terms of inequality. However, according to the Pietra index, the
      society with income distribution $F_a$ is less unequal than the
      society with income distribution $F_b$.
	
      In the second example, two Lorenz functions were considered of
      which the first one is the standard uniform distribution defined
      on any compact interval of the form $[0,b]$ with $b>0$, that is,
      $L_{F_{\bar{u}}}(p)=p^2$ for all $p\in [0,1]$. The other Lorenz
      function has the following form:
        
	\begin{equation*}
	L_{F_S}(p)=
	\left\{ \begin{array}{ll}
	p^2  & \mbox{if $p\in [0,3/4]$,} \\
	1-\frac{7(1-p)}{4}  & \mbox{if $p\in (3/4,1]$.} 
	\end{array}
	\right.  
      \end{equation*}

      $\mathcal{K}_{F_{\bar{u}}} = \mathcal{K}_{F_S} = 2/\phi - 1 <
      \mathcal{G}_{F_S} = 21/64 < \mathcal{G}_{F_{\bar{u}}} =
      1/3$. This example demonstrates an important difference between
      $\mathcal{K}_F$ and $\mathcal{G}_F$. The Gini index is affected
      by transfers within a group. In particular, the poor are
      unaffected but the rich (lying in the interval $[3/4,1)$) have
      become more egalitarian while moving from $L_{F_{\bar{u}}}$ to
      $L_{F_S}$. The normalized $k$-index on the other hand is
      unaffected with such intra-group transfers. Therefore, if we are
      interested in reducing inequality between groups, then the
      normalized $k$-index is a better indicator than the Gini index.
      
      \section{Numerical observations}
      \noindent For the purpose of comparison between different inequality
      indices, we present in Table~\ref{tab:1}, the
        estimated values of the Gini and $k$-indices for the income
        distributions in some countries for the period 1963-1983.
        Tables~\ref{tab:2} and \ref{tab:3} give the estimated values
        of these indices along with the Pietra index for citations,
        for different institutions and universities across the world
        observed in different years.  Table~\ref{tab:4} also shows the
        comparison between Gini, Pietra and $k$ for inequalities in
        paper citations for various science journals.  All the tables
      are taken from \cite{GC2014}.

\begin{figure}[H]
   \centering
  	\includegraphics[width=8cm]{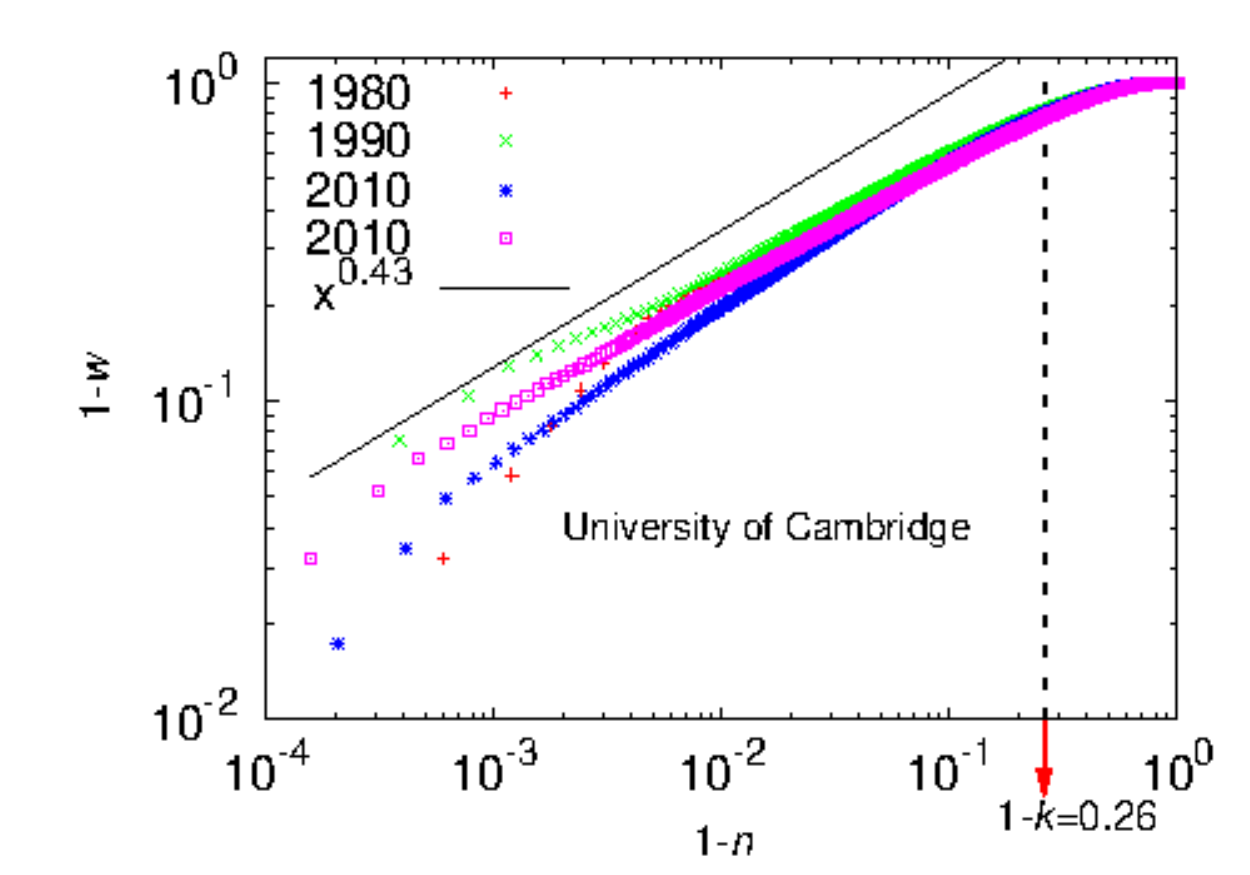}
  	\includegraphics[width=8cm]{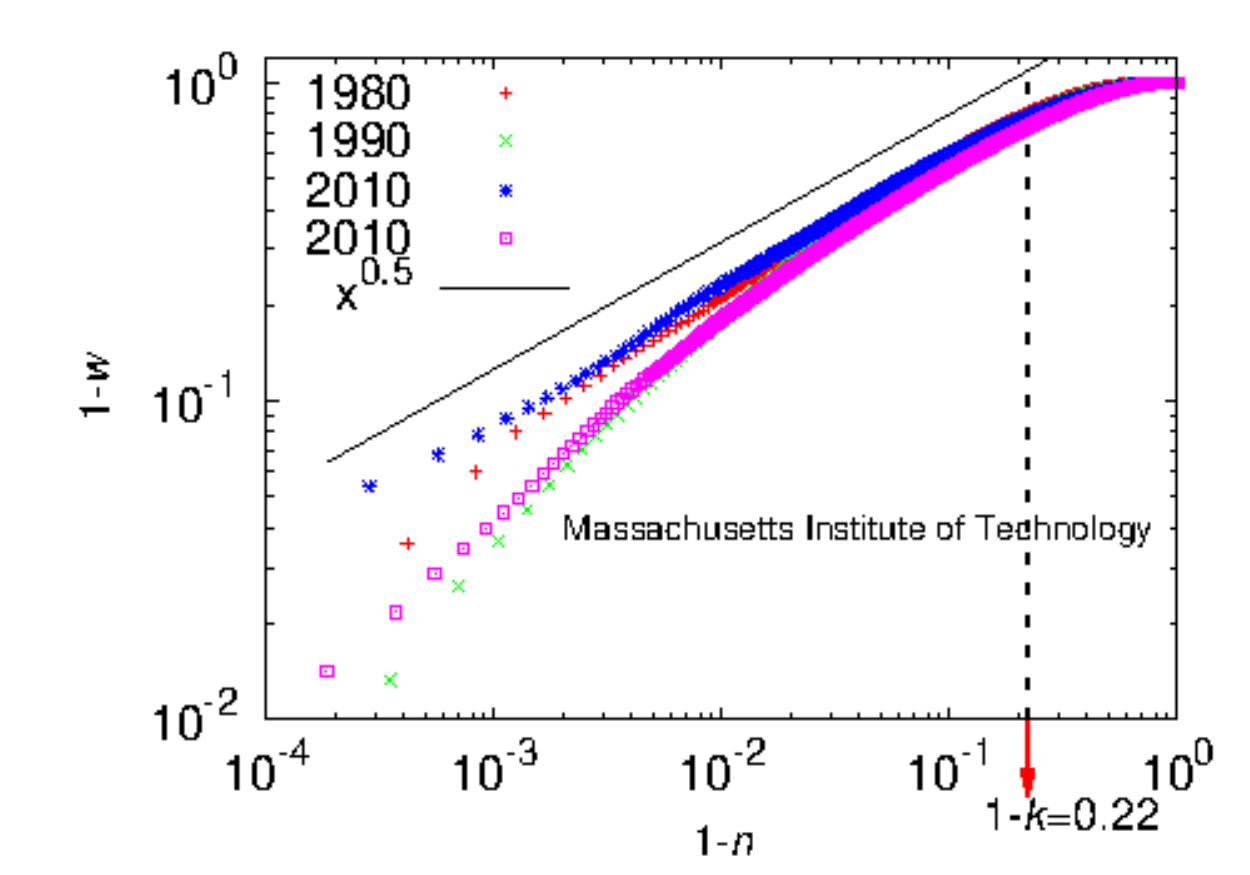}
  	\caption{Illustration of the power law in the citation
          distributions for Cambridge and MIT.  Here,
          $1-w\sim(1-n)^\alpha$ for $n\geq k$, with
          $\alpha=0.50\pm0.10$ [Adapted from \cite{GC2014}].}
  	\label{power-law}
  \end{figure}

\begin{figure}[H]
   \centering
  	\includegraphics[width=8cm]{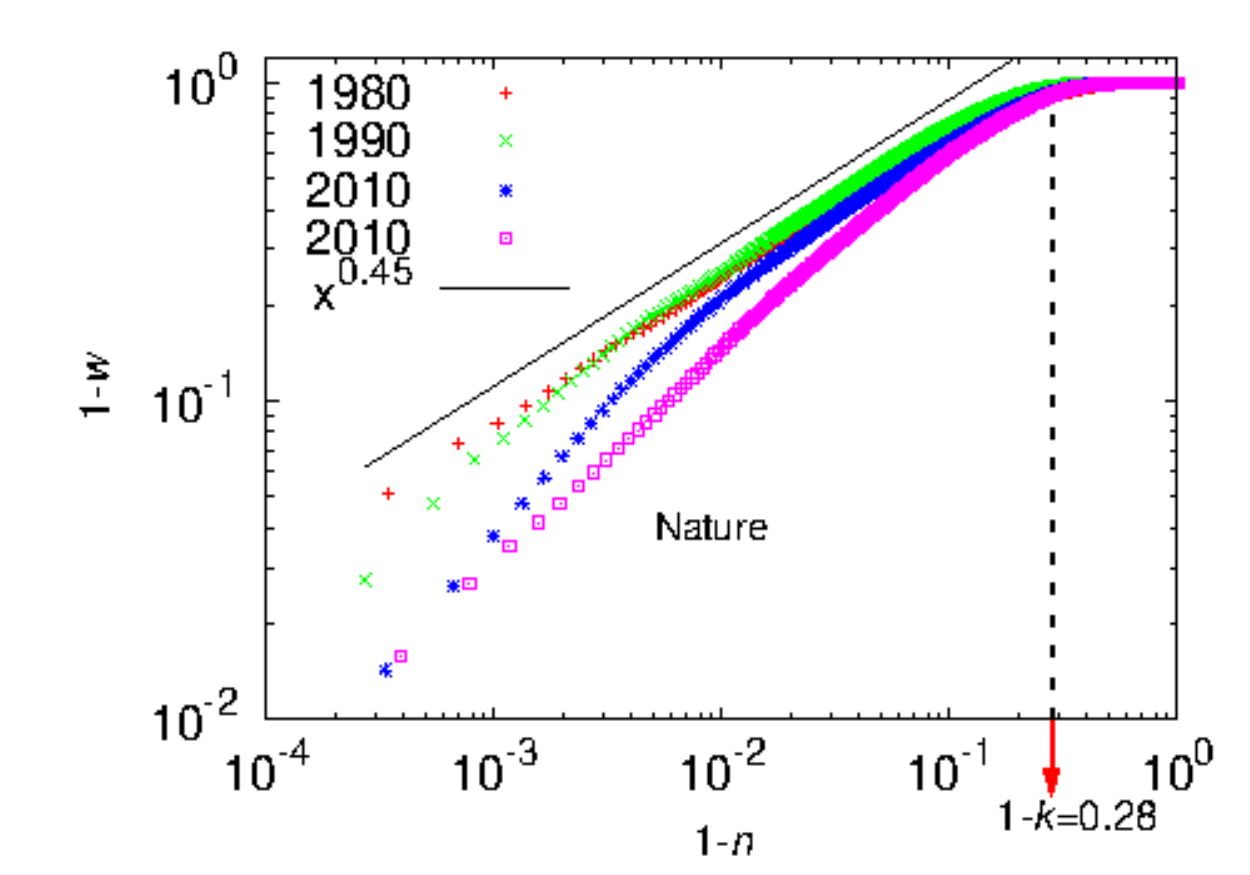}
  	\includegraphics[width=8cm]{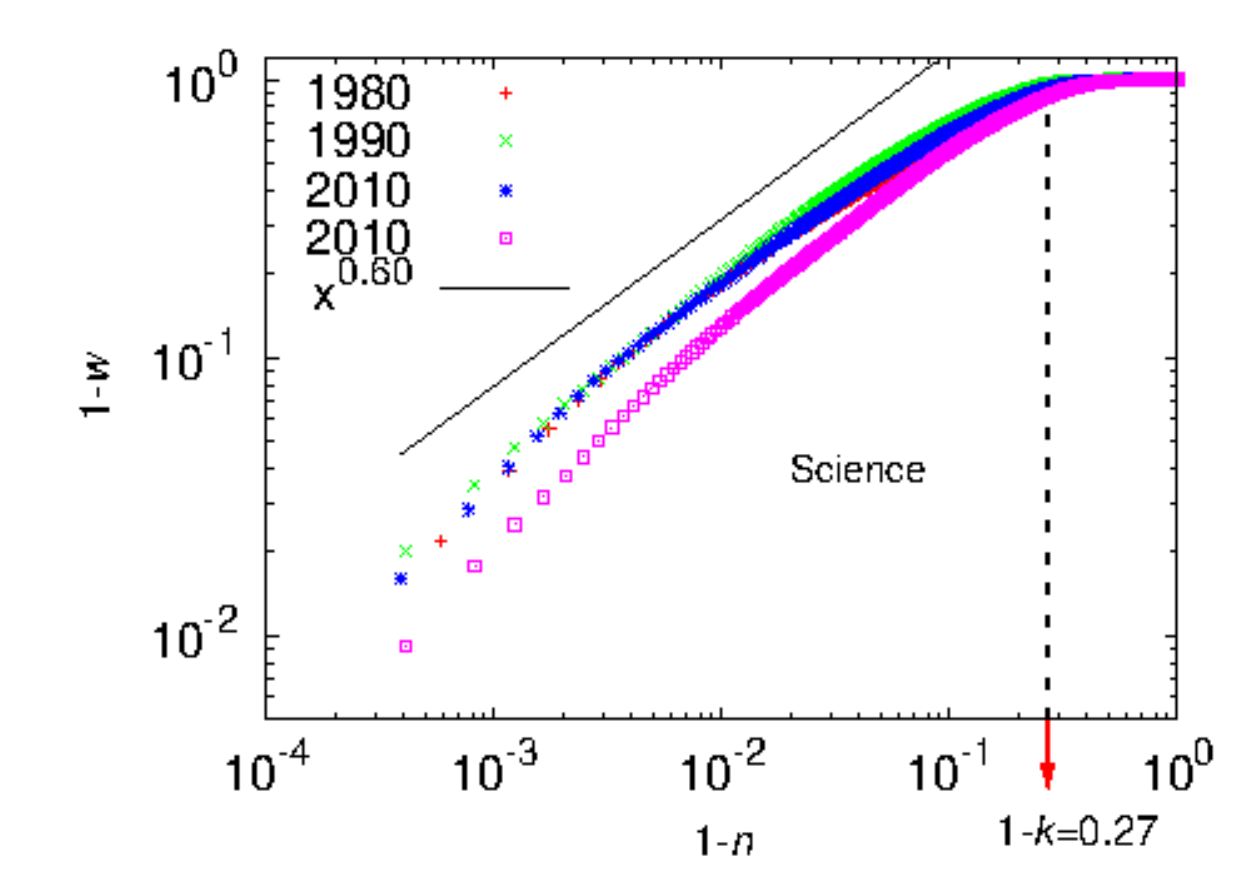}
  	\caption{Illustration of the power law in the citation
          distributions for \emph{Nature} and \emph{Science}. Here,
         $1-w\sim(1-n)^\alpha$ for $n\geq k$,
          with $\alpha=0.50\pm0.10$ [Adapted from \cite{GC2014}].}
  	\label{power-law2}
  \end{figure}

      In \cite{GC2014} it was observed that equation (\ref{eq:gkrel})
      is an approximate result and can differ for large values of $G$
      and $k$. Furthermore, the value of $k$ corresponds to an upper
      limit beyond which the distribution follows a power law pattern,
      similar to the celebrated Pareto law \cite{P1964}.  For the
      inequality in citation data, if $n$ is the fraction of papers
      and $w$ is the cumulative fraction of citations, then for
      $n \geq k$, $1-w\sim (1-n)^\alpha$ with $\alpha=0.50\pm 0.10$
      which implies $n=1-c*(1-w)^\nu$ for $\nu=2.0\pm 0.5$ and $c$ is
      a proportionality constant.  This is illustrated in Figures
      \ref{power-law} and \ref{power-law2}.

\begin{table}[H]
    \centering
    \begin{tabular}{|l|c|c|}
  	\hline                        
  	Country & Gini index & k-index \\
  	\hline 
  	Brazil & 0.62 & 0.73 \\
  	\hline 
  	Denmark & 0.36 & 0.63 \\
  	\hline  
  	India & 0.45 & 0.66 \\
  	\hline 
  	Japan & 0.31 & 0.61 \\
  	\hline  
  	Malaysia & 0.50 & 0.68 \\
  	\hline 
  	New Zealand & 0.37 & 0.63 \\
  	\hline  
  	Panama & 0.44 & 0.66 \\
  	\hline 
  	Sweden & 0.38 & 0.64 \\
  	\hline 
  	Tunisia & 0.50 & 0.69 \\
  	\hline 
  	Uruguay & 0.49 & 0.68 \\
  	\hline  
  	Columbia & 0.55 & 0.70 \\
  	\hline 
  	Finland & 0.47 & 0.67 \\
  	\hline 
  	Indonesia & 0.44 & 0.65 \\
  	\hline
  	Kenya & 0.61 & 0.73 \\
  	\hline  
    	Netherlands & 0.44 & 0.66 \\
      \hline
      Norway & 0.36 & 0.63 \\
  	\hline  
  	Sri Lanka & 0.40 & 0.65 \\
  	\hline  
  	Tanzania & 0.53 & 0.70 \\
  	\hline 
  	United Kingdom & 0.36 & 0.63 \\
  	\hline 
  	Australia & 0.34 & 0.62 \\
  	\hline 
  	Canada & 0.34 & 0.62 \\
  	\hline 
  	Netherlands & 0.31 & 0.61 \\
  	\hline
  	Norway & 0.31 & 0.61 \\
  	\hline  
  	Sweden & 0.29 & 0.60 \\
  	\hline 
  	Switzerland & 0.38 & 0.63 \\
  	\hline 
  	Germany & 0.31 & 0.61 \\
  	\hline 
  	United Kingdom & 0.34 & 0.62 \\
  	\hline  
  	United States & 0.36 & 0.63 \\
  	\hline 
  	\end{tabular}
  	%\vskip1ex
        \caption{The Gini and $k$-indices for the income
          distributions of various countries, $1963\text{--}1983$. The maximum
          error bar in estimated Gini and $k$ values is $\simeq 0.01$
          [Adapted from \cite{GC2014}].}
    \label{tab:1}
\end{table}

\begin{table}[H]
    \centering
    \begin{tabular}{|l|c|c|c|c|c|c|}
           \hline
           Inst./Univ. & Year & Total papers & Citations & Gini index & Pietra index & k-index \\
           \hline 
                       & 1980 & 866 & 16107 & 0.67 & 0.51 & 0.75 \\
  		
  		Melbourne & 1990 & 1131 & 30349 & 0.68 & 0.50 & 0.75 \\
  		
  		& 2000 & 2116 & 57871 & 0.65 & 0.49 & 0.74 \\
  		
  		& 2010 & 5255 & 63151 & 0.68 & 0.50 & 0.75 \\
  		\hline 
  		& 1980 & 2871 & 60682 & 0.69 & 0.52 & 0.76 \\
  		Tokyo     & 1990 & 4196 & 108127 & 0.68 & 0.51 & 0.76 \\
  		& 2000 & 7955 & 221323 &  0.70 & 0.53 & 0.76 \\
  		& 2010 & 9154 & 91349 & 0.70 & 0.52 & 0.76 \\
  		\hline 
  		& 1980 & 4897 & 225626 & 0.73 & 0.55 & 0.78 \\
  		Harvard   & 1990 & 6036 & 387244 & 0.73 & 0.55 & 0.78 \\
  		& 2000 & 9566 & 571666 & 0.71 & 0.54 & 0.77 \\
  		& 2010 & 15079 & 263600 & 0.69 & 0.52 & 0.76 \\
  		\hline 
  		& 1980 & 2414 & 101929 & 0.76 & 0.59 & 0.79 \\
  		MIT    & 1990 & 2873 & 156707 & 0.73 & 0.56 & 0.78 \\
  		& 2000 & 3532 & 206165 & 0.74 & 0.56 & 0.78 \\
  		& 2010 & 5470 & 109995 & 0.69 & 0.51 & 0.76 \\
  		\hline 
  		& 1980 & 1678 & 62981 & 0.74 & 0.56 & 0.78 \\
  		Cambridge & 1990 & 2616 & 111818 & 0.74 & 0.56 & 0.78 \\
  		& 2000 & 4899 & 196250 & 0.71 & 0.54 & 0.77 \\
  		& 2010 & 6443 & 108864 & 0.70 & 0.52 & 0.76 \\
  		\hline 
  		& 1980 & 1241 & 39392 & 0.70 & 0.53 & 0.77 \\
  		Oxford  & 1990 & 2147 & 83937 & 0.73 & 0.56 & 0.78 \\
  		& 2000 & 4073 & 191096 & 0.72 & 0.54 & 0.77 \\
  		& 2010 & 6863 & 114657 & 0.71 & 0.53 & 0.76 \\
  		\hline 
  	\end{tabular}
  	        \caption{The Gini coefficient, Pietra and $k$-indices for
          citations (up to December 2013) of the papers published from
          different universities as obtained from ISI web of
          science. The number of papers and citations give an idea of
          the data size involved in the analysis. [Adapted from
          \cite{GC2014},\cite{CG2017}].}
    \label{tab:2}
\end{table}

\begin{table}[H]
    \centering
    \begin{tabular}{|c|c|c|c|c|c|c|}
  		\hline
  		Inst./Univ. & Year & Total papers & Citations & Gini index & Pietra index & k-index \\
  		\hline 
  		& 1980 & 32 & 170 & 0.72 & 0.49 & 0.74 \\
  		
  		SINP      & 1990 & 91 & 666 & 0.66 & 0.47 & 0.73 \\
  		
  		& 2000 & 148 & 2225 & 0.77 & 0.57 & 0.79 \\
  		
  		& 2010 & 238 & 1896 & 0.71 & 0.52 & 0.76 \\
  		\hline 
  		& 1980 & 450 & 4728 & 0.73 & 0.56 & 0.78 \\
  		IISC      & 1990 & 573 & 8410 & 0.70 & 0.53 & 0.76 \\
  		& 2000 & 874 & 19 167 &  0.67 & 0.50 & 0.75 \\
  		& 2010 & 1624 & 11 497 & 0.62 & 0.45 & 0.73 \\
  		\hline 
  		
  		& 1980 & 167 & 2024 & 0.70 & 0.52 & 0.76 \\
  		TIFR      & 1990 & 303 & 4961 & 0.73 & 0.54 & 0.77 \\
  		& 2000 & 439 & 11 275 & 0.74 & 0.55 & 0.77 \\
  		& 2010 & 573 & 9988 & 0.78 & 0.59 & 0.79 \\
  		\hline 
  		& 1980 & 162 & 749 & 0.74 & 0.56 & 0.78 \\
  		Calcutta    & 1990 & 217 & 1511 & 0.64 & 0.48 & 0.74 \\
  		& 2000 & 173 & 2073 & 0.68 & 0.50 & 0.74 \\
  		& 2010 & 432 & 2470 & 0.61 & 0.45 & 0.73 \\
  		\hline 
  		& 1980 & 426 & 2614 & 0.67 & 0.49 & 0.75 \\
  		Delhi   & 1990 & 247 & 2252 & 0.68 & 0.52 & 0.76 \\
  		& 2000 & 301 & 3791 & 0.68 & 0.51 & 0.76 \\
  		& 2010 & 914 & 6896 & 0.66 & 0.49 & 0.74 \\
  		\hline 
  		& 1980 & 193 & 1317 & 0.69 & 0.53 & 0.76 \\
  		Madras   & 1990 & 158 & 1044 & 0.68 & 0.52 & 0.76 \\
  		& 2000 & 188 & 2177 & 0.64 & 0.47 & 0.73 \\
  		& 2010 & 348 & 2268 & 0.78 & 0.58 & 0.79 \\
  		\hline 
  		
  	\end{tabular}
  	\vskip1ex
        \caption{The Gini, Pietra and $k$-indices for citations (up to
          December 2013) of the papers published from different Indian
          universities, as obtained from ISI web of science [Adapted
          from \cite{GC2014}].}
    \label{tab:3}
\end{table}
\begin{table}[H]
    \centering
    \begin{tabular}{|c|c|c|c|c|c|c|}
  	\hline
  	Journals & Year & Total papers & citations & Gini index & Pietra index & k-index \\
  	\hline 
  	& 1980 & 2904 & 178927 & 0.80 & 0.63 & 0.81 \\
  	
  	Nature    & 1990 & 3676 & 307545 & 0.86 & 0.72 & 0.85 \\
  	
  	& 2000 & 3021 & 393521 & 0.81 & 0.65 & 0.82 \\
  	
  	& 2010 & 2577 & 100808 & 0.79 & 0.63 & 0.81\\
  	\hline 
  	& 1980 & 1722 & 111737 & 0.77 & 0.60 & 0.80 \\
  	Science   & 1990 & 2449 & 228121 & 0.84 & 0.70 & 0.84 \\
  	& 2000 & 2590 & 301093 & 0.81 & 0.66 & 0.82\\
  	& 2010 & 2439 & 85879 & 0.76 & 0.60 & 0.79 \\
  	\hline 
  	
  	& 1980 & - & - & - & - \\
  	PNAS(USA) & 1990 & 2133 & 282930 & 0.54 & 0.39 & 0.70 \\
  	& 2000 & 2698 & 315684 & 0.49 & 0.35 & 0.68 \\
  	& 2010 & 4218 & 116037 & 0.46 & 0.33 & 0.66 \\
  	\hline 
  	& 1980 & 394 & 72676 & 0.54 & 0.39 & 0.70 \\
  	Cell      & 1990 & 516 & 169868 & 0.50 & 0.36 & 0.68 \\
  	& 2000 & 351 & 110602 & 0.56 & 0.41 & 0.70 \\
  	& 2010 & 573 & 32485 & 0.68 & 0.51 & 0.75 \\
  	\hline 
  	& 1980 & 1196 & 87773 & 0.66 & 0.48 & 0.74 \\
  	PRL     & 1990 & 1904 & 156722 & 0.63 & 0.47 & 0.74 \\
  	& 2000 & 3124 & 225591 & 0.59 & 0.43 & 0.72 \\
  	& 2010 & 3350 & 73917 & 0.51 & 0.37 & 0.68 \\
  	\hline 
  	& 1980 & 639 & 24802 & 0.61 & 0.45 & 0.73 \\
  	PRA      & 1990 & 1922 & 54511 & 0.61 & 0.45 & 0.72 \\
  	& 2000 & 1410 & 38948 & 0.60 & 0.44 & 0.72 \\
  	& 2010 & 2934 & 26314 & 0.53 & 0.38 & 0.69 \\
  	\hline 
  	& 1980 & 1413 & 62741 & 0.65 & 0.49 & 0.74 \\
  	PRB      & 1990 & 3488 & 153 521 & 0.65 & 0.48 & 0.74 \\
  	& 2000 & 4814 & 155172 & 0.59 & 0.44 & 0.72 \\
  	& 2010 & 6207 & 70612 & 0.53 & 0.38 & 0.69 \\
  	\hline
  	
  	& 1980 & 630 & 19373 & 0.66 & 0.49 & 0.75 \\
  	PRC      & 1990 & 728 & 15312 & 0.63 & 0.46 & 0.73 \\
  	& 2000 & 856 & 19143 & 0.57 & 0.42 & 0.71 \\
  	& 2010 & 1061 & 11764 & 0.56 & 0.40 & 0.70 \\
  	\hline
  	
  	& 1980 & 800 & 36263 & 0.76 & 0.59 & 0.80 \\
  	PRD      & 1990 & 1049 & 33257 & 0.68 & 0.52 & 0.76 \\
  	& 2000 & 2061 & 66408 & 0.61 & 0.45 & 0.73 \\
  	& 2010 & 3012 & 40167 & 0.54 & 0.39 & 0.69 \\
  	\hline
  	
  	& 1980 & - & - & - & - & -\\
  	PRE      & 1990 & - & - & - & - & - \\
  	& 2000 & 2078 & 51860 & 0.58 & 0.42 & 0.71 \\
  	& 2010 & 2381 & 16605 & 0.50 & 0.36 & 0.68 \\
  	\hline
  \end{tabular}
  \vskip1ex
  \caption{The Gini, Pietra and $k$-indices for citations (up to
    December 2013) of the papers published from different journals, as
    obtained from ISI web of science [Adapted from \cite{GC2014}].}
    \label{tab:4}
\end{table}

  \section{Summary and discussions}
  \noindent For the nonlinear Lorenz function $(L_F(p))$,
  the traditional measures like Gini index measures some ``average
  property'', while the Kolkata index ($k$) identifies the non-trivial
  fixed point of the complementary Lorenz function
  ($\hat{L}_F(p)$ = $1- L(p)$; note that $L_F(p)$
      has trivial fixed points at $p = 0$ and $1$, while
      $\hat{L}_F(p)$ has a nontrivial fixed point at $p = k$). This
  $k$-index apart from capturing the essential character of the
  nonlinear Lorenz function (as inspired by the major developments of
  renormalization group theory in statistical physics
  \cite{Fisher1998} or in identifying the universal characters
  corresponding to the onset of chaos in nonlinear systems
  \cite{Feigenbaum1983}), also gives us a very tangible one, giving
  that $(1-k)$ fraction of the population possess $k$ fraction of the
  total wealth in the society. In \cite{Su2010} the $k$-index is used to define a generalized Gini index. In a recent study, the $k$-index has been used to quantify the inequality for spreading of the Covid-19 infection in urban neighbourhoods and slums in a society (see \cite{SJ2020}).
  
      After a general introduction in Section 1, we
      discuss in Section 2, some structural features of the Lorenz
      function and introduce the Complementary Lorenz function, which
      has a nontrivial fixed point (namely the Kolkata index) as
      mentioned above. In Sections 3 and 4, we try to demonstrate the
      uniqueness of the k-index, compared to Gini and Pietra indices
      in ranking the rich-poor disparity, assuming some typical income
      distributions. we have argued (in Section
      $3$) that the procedure of obtaining the $h$-index of any
  research scientist using the generated citation curve is the same as
  identifying the fixed point of the complementary Lorenz function of
  any income distribution that yields the $k$ index. While comparing
  the normalized $k$-index with the Pietra index and with the Gini
  index, one can show that for any given distribution the normalized
  $k$-index is no more than the Pietra index and the Pietra index is
  no more than the Gini index. We have also argued
  (in Section $4.2$) that for any given
  distribution the normalized $k$-index, the Pietra index and the Gini
  index coincide only if either the society is such that all agents
  have equal income or there are only two income groups in a society
  with some added restrictions (see condition C2 in this
  subsection). We have also argued (in Section
      $5$) that if we are interested in reducing inequality between
  the rich and poor groups of the society, then the normalized
  $k$-index is a better indicator than the Gini index.
  In section $6$, we can see that while the Gini
      index value typically ranges from $0.30$ to $0.62$, the Kolkata
      index value ranges from $0.60$ to $0.73$ at any particular time
      or year for income or wealth data across the countries of the
      world.  It may be mentioned here that income inequality data are
      not easily available from reliable sources. On the other hand,
      the (paper) citations may be considered as a measure of the
      wealth created by the respective University or Institution and
      the resulting inequality data are abundantly available in
      accurate digital formats (say from the ISI Web of Science). We
      estimated the Gini, Pietra and Kolkata index values for the
      citations earned by the yearly publications of various academic
      institutions from such data sources.  We find that while Gini
      and Pietra index values range from $0.65$ to $0.75$ and $0.50$
      to $0.60$ respectively, the Kolkata index remains around
      $0.75 \pm 0.05$ value for Institions or Universities across the
      world. As mentioned already, $k$-index is the social equivalent
      to the $h$-index for an individual researcher or academician.
  Also we find that the value for $k$-index gives an estimate of the
  crossover point beyond which the growth of income (or citations)
  with the fraction of population (or publications) enters a power law
  (Pareto) region (see figure \ref{power-law} and \ref{power-law2}).
  
  \section{Appendices}
  
  \subsection{Appendix A}

  We formally show that for the discrete random variable $F_G$ with
  the Lorenz function is given by (\ref{examplezeroLorenz}), the Gini
  index has the following explicit form:
  
  $$\mathcal{G}_{F_G}=\frac{\sum\limits_{g=1}^G\sum\limits_{t=1}^{G}n_tn_g|x_t-x_g|}{2NM}.$$
  
  Observe first that
  
 \begin{equation}\label{step}
 \begin{split}
 \int\limits_{0}^1L_{F_G}(q)dq & = \sum\limits_{g=1}^G\left\{\int\limits_{N(g-1)}^{N(g)}L_{F_G}(p_k)dp_k\right\} \\
 & =\sum\limits_{g=1}^G\left\{\int\limits_{N(g-1)}^{N(g)}\left\{M(g-1)+\left(p_g-N(g-1)\right)\left(\frac{Nx_g}{M}\right)\right\}dp_g\right\}\\ & =-\frac{\sum\limits_{g=1}^G\sum\limits_{t=1}^{g-1}n_gn_t(x_g-x_t)}{NM}+\frac{\sum\limits_{g=1}^G\left(2\sum\limits_{t=1}^{g-1}n_t+n_g\right)n_tx_g}{2NM}.
 \end{split}
 \end{equation}
 
 Thus, using $2\sum\limits_{g=1}^G\left(\sum\limits_{t=1}^{g-1}n_t-\sum\limits_{t=g+1}^Gn_t\right)n_gx_g=\sum\limits_{g=1}^G\sum\limits_{t=1}^{G}n_gn_t|x_g-x_t|$ and using (\ref{step}) we get  
 \begin{equation} \label{step1}
 \begin{split}
 \mathcal{G}_{F_G} & =1-2\int\limits_{0}^1L_{F_G}(q)dq  \\ & =1-\frac{\sum\limits_{g=1}^G\left(2\sum\limits_{t=1}^{g-1}n_t+n_g\right)n_gx_g}{NM}+\frac{2\left\{\sum\limits_{g=1}^G\sum\limits_{t=1}^{g-1}n_gn_t(x_g-x_t)\right\}}{NM} \\ & =\frac{\sum\limits_{g=1}^G\left(\sum\limits_{t=g+1}^Gn_t-\sum\limits_{t=1}^{g-1}n_t\right)n_gx_g}{NM}+\frac{\sum\limits_{g=1}^G\sum\limits_{t=1}^{G}n_gn_t|x_g-x_t|}{NM} \\ & =-\frac{\sum\limits_{g=1}^G\sum\limits_{t=1}^{G}n_gn_t|x_g-x_t|}{2NM}+\frac{\sum\limits_{g=1}^G\sum\limits_{t=1}^{G}n_gn_t|x_g-x_t|}{NM}  \\ &
 =\frac{\sum\limits_{g=1}^G\sum\limits_{t=1}^{G}n_gn_t|x_g-x_t|}{2NM}.
 \end{split}
 \end{equation}

 \noindent
 Hence, from the last inequality in (\ref{step1}) the result follows. 
 
 \subsection{Appendix B} 
 \subsubsection{Appendix B(i).}

 The following derivation shows why $\mathcal{P}_{F}=E(|x-\mu|)/2\mu$
 this is true.
 
 \begin{equation}\label{stepPIdeviation}
 \begin{split}
 \mathcal{P}_{F} & =F(\mu)-L_{F}(\mu)\\ & =F(\mu)-\frac{\int\limits_{0}^{F(\mu)}F^{-1}(q)dq}{\mu} \\
 & =\frac{\int\limits_{0}^{F(\mu)}\left\{\mu-F^{-1}(q)\right\}dq}{\mu}=\frac{2\int\limits_{0}^{F(\mu)}\left\{\mu-F^{-1}(q)\right\}dq}{2\mu}\\ & =\frac{\int\limits_{0}^{F(\mu)}\left\{\mu-F^{-1}(q)\right\}dq+\int\limits_{F(\mu)}^1\left\{F^{-1}(q)-\mu\right\}dq}{2\mu} \\ & =\frac{\int\limits_{0}^{1}|F^{-1}(q)-\mu|dq}{2\mu}\\ & =\frac{E(|x-\mu|)}{2\mu}.
 \end{split}
 \end{equation}
 
 \subsubsection{Appendix B(ii).}

 We formally show that for the discrete random variable $F_G$ with the
 Lorenz function is given by (\ref{examplezeroLorenz}), the Pietra
 index has the following explicit form:
$$\mathcal{P}_{F_G}=\frac{\sum\limits_{g=1}^{\tilde{g}}n_g\left(\mu_G-x_g\right)}{M}=\frac{E(|x-\mu_G|)}{2\mu_G}$$where $\tilde{g}\in \{1,\ldots,G-1\}$ is such that $\mu_G\in [x_{\tilde{g}},x_{\tilde{g}+1})$ implying that $F_G(\mu_G)=N(\tilde{g})$. 
 
For the first equality, observe that there exists $\tilde{g}\in \{1,\ldots,G-1\}$ such that $\mu_G\in [x_{\tilde{g}},x_{\tilde{g}+1})$ implying that $F_G(\mu_G)=N(\tilde{g})$. Thus, using $\sum_{g=1}^Gn_g\left(x_g-\mu_G\right)=0$ and using $F_G(\mu_G)-N(\tilde{g})=0$ we get  

\begin{equation}\label{stepPI}
 \begin{split}
 \mathcal{P}_{F_G} & =F_G(\mu_G)-L_{F_G}(\mu_G)\\ & =F_G(\mu_G)-M(\tilde{g}-1)-\left\{(F_G(\mu_G)-N(\tilde{g}-1))\right\}\left(\frac{Nx_{\tilde{g}}}{M}\right) \\
 & =F_G(\mu_G)\left(\frac{M-Nx_{\tilde{g}}}{M}\right)-\left\{M(\tilde{g}-1)-N(\tilde{g}-1)\left(\frac{Nx_{\tilde{g}}}{M}\right)\right\}\\ & =\frac{F_G(\mu_G)\left(\sum\limits_{g=1}^Gn_g(x_g-x_{\tilde{g}})\right)}{M}+\frac{\sum\limits_{g=1}^{\tilde{g}}n_g(x_{\tilde{g}}-x_g)}{M}\\ & =\frac{\sum\limits_{g=1}^Gn_g\left(x_g-\mu_G\right)}{M}+\frac{\sum\limits_{g=1}^{\tilde{g}}n_g\left(\mu_G-x_g\right)}{M}+\frac{(F_G(\mu_G)-N(\tilde{g}))N(\mu_G-x_{\tilde{g}})}{M}\\ & =\frac{\sum\limits_{g=1}^{\tilde{g}}n_g\left(\mu_G-x_g\right)}{M}.
 \end{split}
\end{equation}

Given (\ref{stepPI}) it follows that the Pietra index of the
distribution $F_G$ with
$\mu_G\in \left[x_{\tilde{g}},x_{\tilde{g}+1}\right)$ is

\begin{equation}\label{oneformPietra}
 \mathcal{P}_{F_G}=\frac{\sum\limits_{g=1}^{\tilde{g}}n_g\left(\mu_G-x_g\right)}{M}. 
 \end{equation}

 Given (\ref{oneformPietra}), we can also derive second equality by
 using $\mu_G\in \left[x_{\tilde{g}},x_{\tilde{g}+1}\right)$ and by
 using
 $\sum_{g=1}^{\tilde{g}}n_g\left(\mu_G-x_g\right)=\sum_{g=\tilde{g}+1}^{n}n_g\left(x_g-\mu_G\right)$. Specifically,

 \begin{equation}\label{stepPIsim}
 \begin{split}
 \mathcal{P}_{F_G} & =\frac{\sum\limits_{g=1}^{\tilde{g}}n_g\left(\mu_G-x_g\right)}{M} \\
 & =\frac{\sum\limits_{g=1}^{\tilde{g}}n_g\left(\mu_G-x_g\right)+\sum\limits_{g=\tilde{g}+1}^{G}n_g\left(x_g-\mu_G\right)}{2M}
 \\ & =\frac{\sum\limits_{g=1}^{G}\left(\frac{n_g}{N}\right)|x_g-\mu_G|}{2\left(\frac{M}{N}\right)}\\ & =\frac{E(|x-\mu_G|)}{2\mu_G}.
 \end{split}
 \end{equation}
  
 \bibliographystyle{unsrt}

\end{document}